\documentclass[12pt]{article}

\usepackage{newtxtext,newtxmath}
\usepackage{graphicx}
\usepackage{tikz}
\usepackage{bm}
\usepackage{comment}

\newcommand{\comments}[1]{}

\usepackage{graphicx}

\usepackage[letterpaper,margin=1in]{geometry}

\usepackage{siunitx}

\renewenvironment{abstract}
	{\quotation}
	{\endquotation}

\date{}

\makeatletter
\renewcommand{\fnum@figure}{\textbf{Figure \thefigure}}
\renewcommand{\fnum@table}{\textbf{Table \thetable}}
\makeatother

\usepackage{scicite}

\usepackage{url}

\def\scititle{
    Glassy Signatures in Water’s Second Liquid
}
\title{\vspace{-2cm}\bfseries \boldmath \scititle}

\author{
	Florian~Pabst$^{1,2\ast}$,
	Ali~Hassanali$^{2\ast}$\and
	\small$^{1}$SISSA – Scuola Internazionale Superiore di Studi Avanzati, 34136 Trieste, Italy\and
	\small$^{2}$CMSP, The Abdus Salam Centre for Theoretical Physics, 34151 Trieste, Italy\and
	\small$^\ast$Corresponding author. Email: fpabst@sissa.it; ahassana@ictp.it
}

\begin{document} 

\maketitle

\begin{abstract} \bfseries \boldmath
The origin of water’s anomalous behavior remains a central open problem in the physical sciences and is often attributed to a liquid–liquid transition (LLT) between high- and low-density liquid states deep in the supercooled regime. Experimental access to this region has been challenging due to rapid crystallization, leaving atomistic simulations as a major source of supporting evidence. Using extensive machine-learning–accelerated first-principles simulations in direct comparison with spectroscopic, structural, and dynamical experimental measurements, we show that features commonly interpreted as signatures of two-liquid behavior coincide with the onset of dramatic dynamical slowing down characteristic of an emerging non-ergodic glassy state. Specifically, we find that two-state fluctuations associated with an LLT, are also consistent with a transformation from a high-density liquid to a kinetically constrained low-density glassy-like state. By mapping equilibrium dynamics across pressure and temperature, our results call for a closer examination of water’s metastable landscape, in which two-state behavior may reflect a relatively high glass-transition temperature of low-density water, 189~$\pm$~8 K---curiously close to the temperature commonly associated with the proposed LLT.
\end{abstract}

\noindent

\clearpage


Cooling makes a liquid denser, less compressible, and less capable of storing thermal energy. Water does the opposite: it expands, and its compressibility and heat capacity rise toward pronounced extrema.
These anomalies have long pointed to a microscopic explanation that, despite decades of intense research, remains unresolved \cite{speedy1976isothermal,kim2017maxima}.
A leading hypothesis attributes them to fluctuations associated with a hidden liquid–liquid critical point (LLCP), buried below the freezing line at elevated pressure. Support for this view emerged from the observation of a first-order–like transition between two amorphous states in experiments \cite{mishima1985apparently}, and later between two liquid states in simulations \cite{poole1992phase}, giving rise to the liquid–liquid transition scenario.

Experimental access to the LLCP has proven extremely challenging because it lies within the crystallization-prone \emph{no-man’s-land}. Only very recent experiments have managed to penetrate this regime, probing various physical quantities on ultra-short time scales before ice formation intervenes \cite{kim2017maxima,kim2020experimental,tyburski2025observation,you2026experimental}. In this regard, molecular dynamics simulations have played an important role in searching for support for the LLCP hypothesis. 

Very recently, atomistic simulations using machine-learning–based potentials trained on first-principles data\cite{gartner2022liquid,sciortino2025constraints} have confirmed the original LLT hypothesis proposed on the basis of simulations with the ST2 potential\cite{poole1992phase}, while studies employing the classical point-charge model TIP4P/2005 have yielded differing conclusions regarding the presence of an LLT\cite{debenedetti2020second,jedrecy2023free,overduin2013analysis,sciortino2024free}. In this respect it has been highlighted recently that the meta-stability between the two liquid states is restricted to a very narrow pressure interval of 200 bars, which may contribute to these conflicting observations\cite{sciortino2024free}. Although these simulations have pushed the boundary to microsecond timescales, comparable to the ultrafast experiments that briefly access deeply supercooled water before crystallization intervenes, equilibration at these low temperatures is known to be challenging due to the very long structural relaxation times. At these conditions, the dynamics becomes glassy, raising the question of whether thermodynamic quantities can still be reliably inferred from simulations. This challenge also translates to experiments deriving thermodynamic support for the presence of an LLCP\cite{you2026experimental}, since they involve measurements limited to the microsecond timescales before the onset of crystallization. But can equilibrium truly be established on these timescales once structural relaxation becomes comparable to, or exceeds, the experimental observation window?

In fact, one of the open questions in the physics of supercooled water is the exact location of its glass transition temperature \cite{amann2016colloquium}. The commonly cited value of 136~K hinges on experiments of hyper-quenched, vapor-deposited or pressure-amorphized samples whose relationship to slowly-supercooled pure water is uncertain. These measurements show only a faint calorimetric signal at this temperature, small enough to be interpreted as a "shadow" transition \cite{yue2004clarifying,giovambattista2004glass}, suggesting that the true liquid–glass transformation may occur at substantially higher temperatures. Calculations based on the facilitation theory demonstrate that the glass transition around 136~K is a non-equilibrium effect imposed by the ultra-fast cooling rates, and it would shift to around 180~K if water could be supercooled with a standard rate on the order of 1~K/min \cite{limmer2014length}. While concentrated salt solutions often exhibit a $T_g$ around 136~K, recent experiments using minimal solute content find a markedly elevated $T_g = 166\pm15$K \cite{lunkenheimer2025exploring}. Furthermore, salts believed not to perturb water structure yield a $T_g$ near 190~K \cite{zhao2016apparent,woutersen2018liquid}. Studies in nanoconfinement---where crystallization is suppressed by geometry---likewise show that the 136~K feature is a shadow glass transition and place the genuine $T_g$ between 170 and 200~K \cite{melillo2024complexity}. 

Very recently, temperature dependent electron diffraction experiments cooling from the liquid at room temperature, have shown that the structure of water does not evolve further below 200~K, suggesting a continuous transition into a hyperquenched glassy state at this temperature for the cooling time of $15$~$\mu$s before measurement\cite{kruger2023electron}. Clearly, there is currently no smoking-gun measurement that unequivocally pinpoints the $T_g$ of slow-cooled bulk water. Most importantly, if water’s $T_g$ indeed lies at the upper end of the range suggested by the aforementioned experiments, it would be very close to the LLT temperature found in the most realistic ab initio–based water models based both on density functional theory (DFT)\cite{zhang2021phase} and many-body polarization (MB-pol)\cite{bore2023realistic}, corrected for their melting point mismatch with respect to experiment. Notably, a recent simulation study of the energy landscape of the MB-pol model reports that the configurational entropy of water vanishes at temperatures much higher than 136~K \cite{szukalo2026energy}. Because an equilibrated liquid cannot persist once its configurational entropy is effectively exhausted, these temperatures suggest a lower bound for $T_g$, in agreement with the experimental observations of $T_g \gg 136$~K discussed above.

These observations raise an intriguing possibility: if deeply supercooled water experiences a marked slowdown of the dynamics around the temperature of the LLCP, could signatures often attributed to a liquid–liquid transition also be consistent with a transition between a fluid high-density liquid (HDL) and a kinetically constrained low-density glassy state? To tackle this question, we explore how the equilibrium dynamics of deeply supercooled water evolve as a function of temperature and pressure, comparing state-of-the-art machine-learning–accelerated first-principles simulations with experimental measurements in regimes where crystallization can be suppressed. This allows us to identify the conditions under which liquid water remains ergodic and when it instead falls out of equilibrium---distinguishing genuine liquid behavior from glassy non-equilibrium states. By establishing the location of the glass-transition boundary relative to the proposed LLT, we suggest that such signatures may not necessarily reflect an equilibrium transformation between two liquids, but could also be consistent with a crossover between a liquid and a glassy state.

\subsection*{Fluctuations between a liquid and a glassy state}

\begin{figure}[h!]
    \centering
    \includegraphics[width=0.6\linewidth]{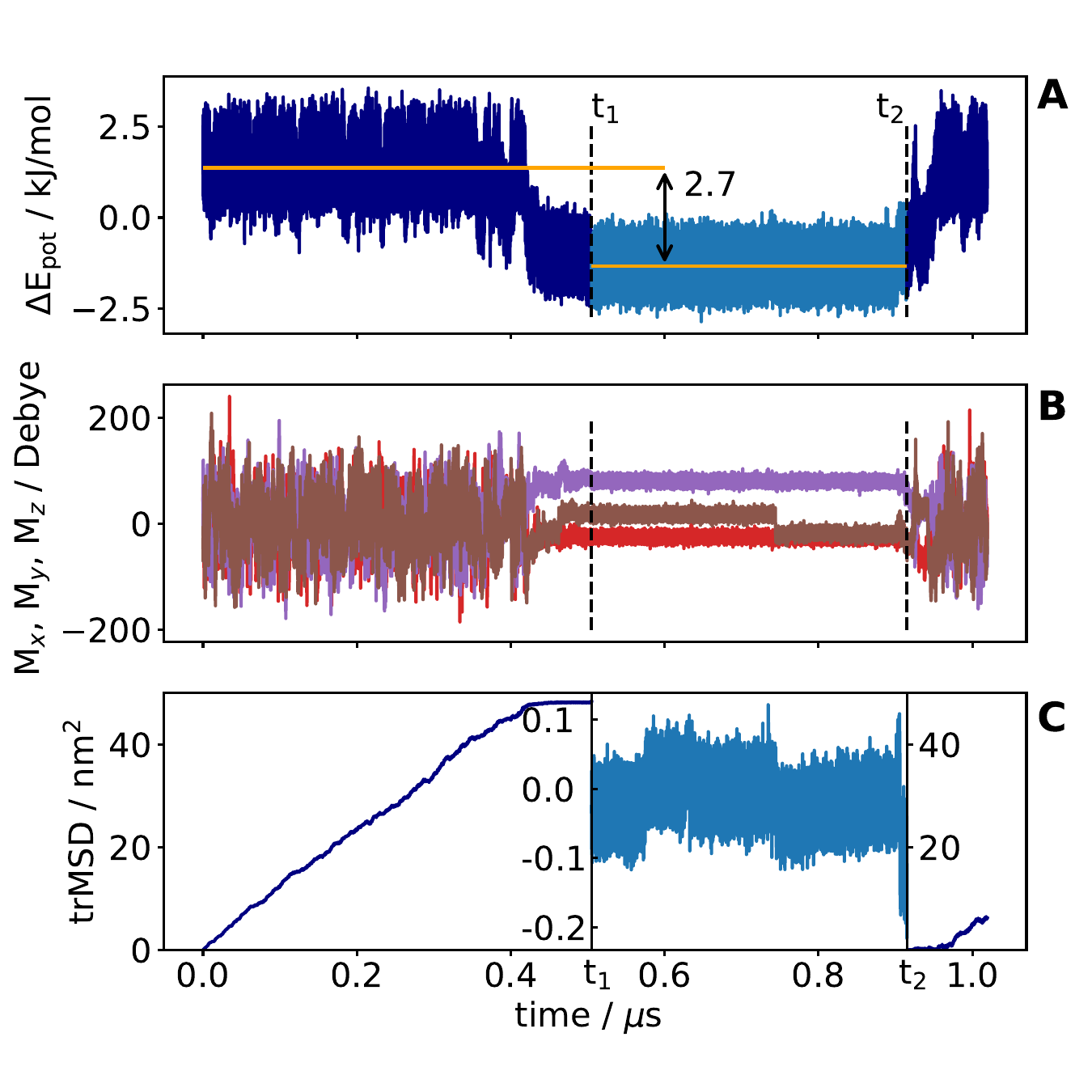}
    \includegraphics[width=0.3\linewidth]{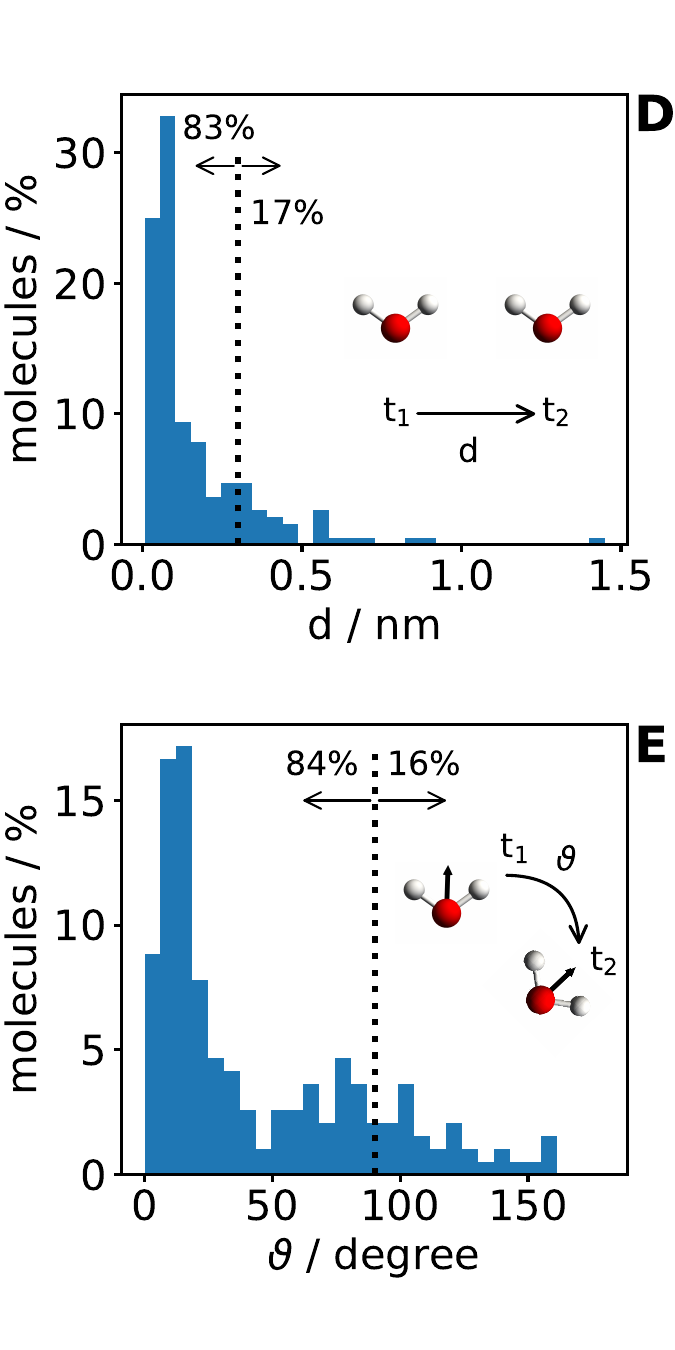}
    \caption{\textbf{Two state fluctuations.} NPT simulation employing the DNN@SCAN potential at 195~K and 3200~bar with 192 water molecules shows fluctuations between a high energy (high density) and a low energy (low density) state (\textbf{A}). The rotational (\textbf{B}) and translational (\textbf{C}) dynamics of the two states characterized by the electric dipole moment in three cartesian directions and the time resolved mean square displacement (trMSD, eq.~\ref{eq:trMSD}), respectively, differ dramatically between the two states: While the system is liquid-like in the high-energy phase, as seen by the dipole components fluctuating rapidly around zero and a near-linear trMSD, the dynamics are strongly reduced in the low-energy phase. This is shown again more quantitatively in panel (\textbf{D}) and (\textbf{E}), where 83\% of the molecules diffuse less than a molecular diameter (0.3~nm) and 84\% rotate less than 90° during the $\sim$0.4~$\mu$s long stay in the low-density state.}
    \label{fig:1}
\end{figure}

To appreciate the extent of the glassy dynamics that develops during the formation of the purported low-density liquid (LDL) phase, we performed a simulation with the state-of-the-art deep neural-network (DNN) potential DNN@SCAN, for the same state point where LDL$\leftrightarrow$HDL fluctuations were previously reported \cite{gartner2022liquid}, i.e., 195~K and 3200~bar (all reported temperatures from DNN simulations in this work are shifted by their mismatch in melting temperature with respect to the experimental value to facilitate comparison between simulation and experiment) and employing 192 water molecules. During the more than 1~$\mu$s long simulation, we focus on illustrating the correlation between energy (panel A Fig.~\ref{fig:1}) and molecular motion (panel B and C) to distinguish liquid behavior from glassy features. 

To probe the collective dynamics of the system, we determined the total electric dipole moment of the system in three cartesian directions, reporting on the collective dynamics of the rotational degrees of freedom (panel B), as well as the time resolved mean square displacement (trMSD) defined as,
\begin{equation}
    \text{trMSD}(t) = \frac{1}{N}\sum\limits_{i=1}^{N} |\mathbf{r}_i(t)-\mathbf{r}_i(0)|^2,
    \label{eq:trMSD}
\end{equation}
which reports on the global translational motion traversed by the molecules as a function of time (panel C). A striking feature observed from this analysis is the presence of a long-lived period where the dipole moment remains essentially flat and the trMSD exhibits only small fluctuations. These periods correspond to a low value of density and potential energy, during which the system is stuck in a dynamically restricted, low-density state characterized by small rotational and translational molecular motions. These small-amplitude motions are quantified in panel D (translation) and panel E (rotation), which show that 83\% of the molecules translate less than one molecular diameter (0.3~nm) and 84\% rotate less than 90°, i.e., the threshold required for decorrelation, during the low-density period of $\sim$0.4~$\mu$s. In sharp contrast, the high-energy (high-density) phase shows substantially enhanced dynamics, represented by the dipole moment fluctuating around zero and a approximately linear increase in trMSD. The coexistence of mobile and immobile regions in the low density state seen here is consistent
with the dynamic heterogeneity well documented in supercooled water and
other glass-forming liquids\cite{malosso2024evidence,ediger2000spatially}. What is striking here is the scale of this
heterogeneity: the low-density state develops a macroscopic collective
polarization that persists for the full $\sim 0.4~\mu\mathrm{s}$,
with a net dipole moment that does not decay to zero. Such a long-lived
collective observable is not what one expects of an equilibrated liquid,
where orientational correlations should average out on times short
relative to the observation window; rather, it points to a breakdown of
ergodicity on these timescales. To illustrate that these features are present also in the low-density states of trajectories from the literature
\cite{gartner2022liquid, sciortino2025constraints, debenedetti2020second},
we repeat the analysis shown in Fig.~\ref{fig:1}A--C for those
data (see Supplementary Information (SI) Fig.~S1) which display similar characteristics. Importantly, the mean polarization of the different low-density segments feature a non-zero mean polarization in different directions (see Fig.~\ref{fig:pol}).

The implications of this breakdown of ergodicity merit attention.
For instance, free-energy landscapes reconstructed from such trajectories
\cite{debenedetti2020second,gartner2022liquid,sciortino2025constraints}
may be sensitive to the non-equilibrium character of the low-energy
phase. Because the system does not fully sample the configurational space
on accessible timescales, particularly the collective orientational
degrees of freedom, the configurational entropy may be underestimated.
This, in turn, can affect the reliable inference of thermodynamic
properties, such as the dielectric constant of the low-density state, a
quantity that should be converged for a true liquid. These
considerations are further highlighted by the sensitivity of the
low-density-state behavior to simulation parameters and finite-size
effects, which can alter the dynamics and effective barriers associated
with transitions and metastability. For example, trapping can be
suppressed by modifying the barostat settings
(see Fig.~\ref{fig:barostat}), while fluctuations involving collective reorientational dynamics are markedly reduced
in larger simulation boxes using DNN@SCAN\cite{malosso2024evidence}.

\subsection*{A Pressure or Temperature Driven Glass Transition}

Identifying how these fluctuations in supercooled water are connected to a glass transition requires directly interrogating the kinetics of glassy dynamics. To this end, we design simulations that independently induce vitrification through changes in temperature or pressure, motivated by two experimental protocols that have been used to infer the existence of distinct liquid phases. The first is inspired by X-ray scattering experiments that track the depressurization dynamics of high-density liquid water following laser-induced heating of high-density amorphous ice (HDA) \cite{kim2020experimental}, while the second mirrors calorimetric measurements upon cooling. For all the ensuing analysis, unless stated otherwise, we report results coming from extensive simulations with the DNN@SCAN potential using 512 water molecules.

\begin{figure}[h!]
\centering
\includegraphics[width=0.49\textwidth]{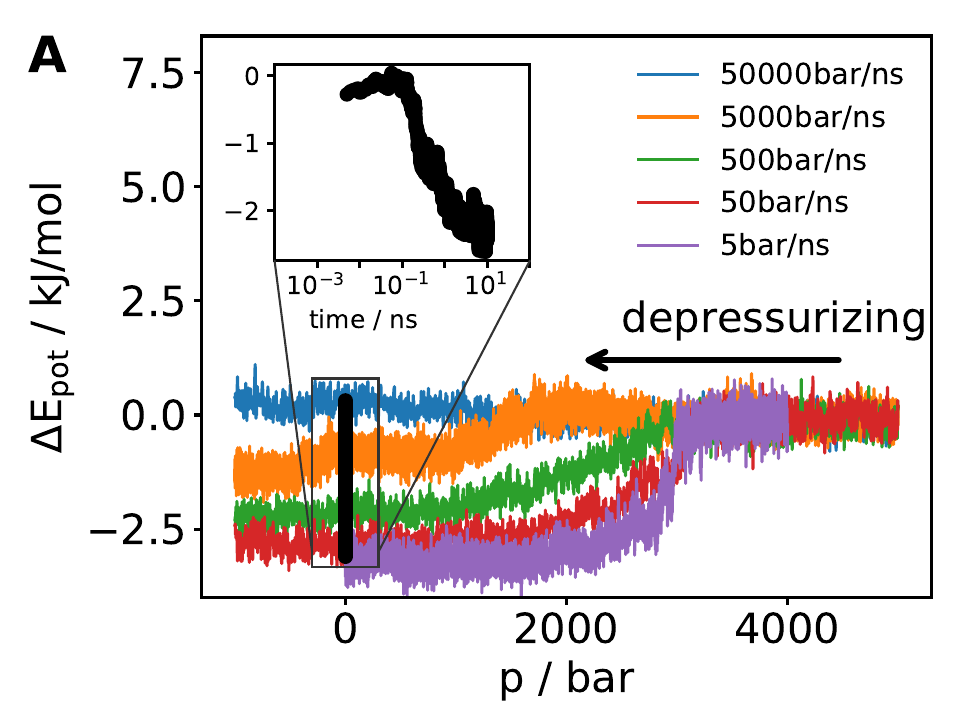}
\includegraphics[width=0.49\textwidth]{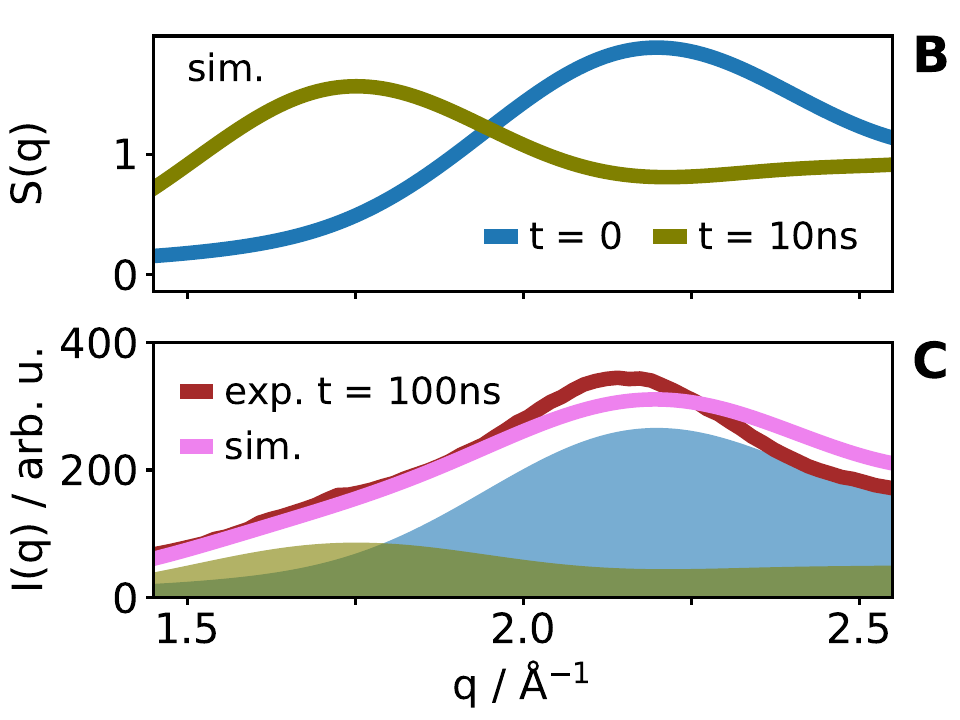}
\includegraphics[width=0.49\textwidth]{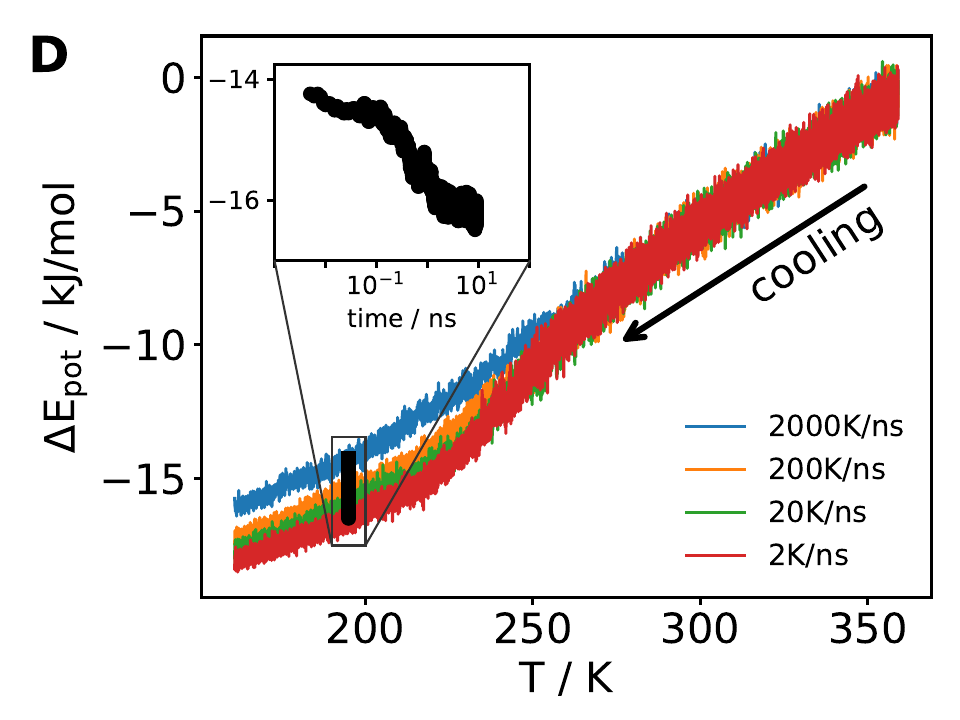}
\includegraphics[width=0.49\textwidth]{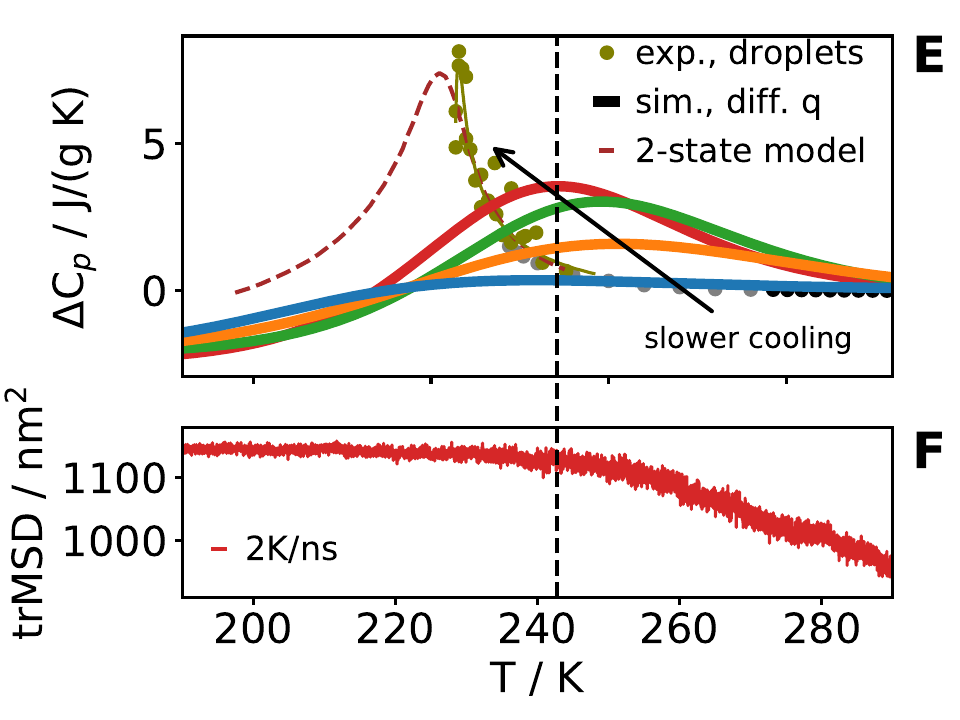}
\caption{\linespread{0.9}\selectfont  \textbf{Pressure- and cooling-driven vitrification.}
(\textbf{A}) Potential energy (in kJ/mol per molecule) during depressurization of equilibrated high-density liquid (HDL) configurations at 195~K using different pressure-ramp rates. Slower ramps produce progressively lower-energy states, characteristic of pressure-induced glass formation upon depressurization. Inset: physical aging after rapid depressurization reflects slow relaxation toward equilibrium.
(\textbf{B,C}) Comparison to laser-heating X-ray experiments \cite{kim2020experimental}. Simulated structure factors before and after depressurization reproduce the experimentally observed low-$q$ shift. The experimental signal is consistent with a mixture of HDL and a low-density amorphous state.
(\textbf{D}) Cooling at ambient pressure shows rate-dependent trapping in lower-energy glassy states; inset: aging after fast cooling matches the energy scale of the apparent phase fluctuations.
(\textbf{E}) Heat capacity during cooling evolves from a step-like glass-transition signature at fast rates to an additional peak at slower rates $q$, approaching experimental droplet data.
(\textbf{F}) Translational dynamics show that slowing dynamics coincide with the $C_p$ maximum, indicating that vitrification cuts off the divergence of $C_p$ on accessible timescales.
For the dynamics during depressurizing see Fig.~\ref{fig:dynArrest}.}
\label{fig:glassTransition}
\end{figure}

\emph{Pressure Induced Vitrification:} Figure~\ref{fig:glassTransition}A shows simulations performed at 195~K corresponding to the temperature where fluctuations between the high-density liquid and the low-density phase were observed in Fig.~\ref{fig:1}. Starting from well-equilibrated HDL configurations at high pressure (5~kbar), we depressurized the system to $-1$~kbar using five different pressure-ramp rates. Except for the fastest ramp, where the potential energy remains essentially unchanged, all slower ramps drive the system into a distinctly lower-energy state. Moreover, the final energy decreases systematically with decreasing ramp rate. This mirrors the classical signature of glass formation: slower driving allows the system to explore deeper minima in the potential-energy landscape. Although pressure-induced glass transitions are less common than temperature-driven ones, they are well documented \cite{shumway1995molecular}; the key difference here is that the transition occurs upon depressurization rather than pressurization, highlighting the unusual thermodynamic landscape of supercooled water.

Importantly, the change in potential energy produced by these depressurization ramps closely matches the energy drop ($\sim$2.7~kJ/mol) observed during the phase fluctuations in Fig.~\ref{fig:1}. This indicates that in those trajectories, the system intermittently becomes \emph{kinetically trapped} in low-energy configurations corresponding to states normally accessible only at lower pressure. In other words, the transitions observed in Fig.~\ref{fig:1} reflect spontaneous trapping and release of the low-density glassy states.

The inset of Fig.~\ref{fig:glassTransition}A shows the isobaric evolution of $\Delta E_{\rm pot}$ at 1~bar after depressurization from 5~kbar with 50~kbar/s. As a consequence of the rapid ramp that drives the system far from equilibrium, the subsequent evolution reflects physical aging toward the equilibrium liquid. Although both the main panel and inset might suggest that equilibration lies within reach at the longest simulated times or slowest ramps, we will see shortly that a rigorous equilibration would require many orders of magnitude longer timescales than that accessible to simulations at these conditions. It is important to emphasize that this kind of non-equilibrium dynamics following a sudden transition (such as the high- to low-density transition in Fig.~\ref{fig:1}) may give the appearance of enhanced mobility, but should not be mistaken for equilibrium dynamics.

Figure~\ref{fig:glassTransition}B and C link our theoretical results directly to X-ray scattering experiments \cite{kim2020experimental}. In those measurements, depressurization following laser-induced heating produces a low-q peak in the structure factor, which was interpreted as the rapid formation of a low-density liquid (LDL). Our simulation protocol---depressurizing HDL at $50$~kbar/ns followed by isobaric relaxation at ambient pressure---closely mirrors this experimental procedure. Figure~\ref{fig:glassTransition}B shows the simulated structure factors at t=0 and after 10 ns at ambient pressure, matching the experimental temporal resolution; over this interval, the principal peak shifts to lower q, consistent with the density decrease accompanying the transformation to a lower-density state. Panel C compares the simulated scattering to the experimental signal measured 100 ns after laser heating. Following the experimental analysis, we fit the experimental data with a superposition of high- and low-density structure factors from panel B, yielding a low-density fraction of 25\% and an HDL fraction of 75\%, in accordance with the populations reported in the experimental study. Crucially, our simulations show that the low-density component identified in this manner is glassy on those timescales rather than an equilibrated liquid, highlighting that the experiments do not directly distinguish between a liquid and a glassy low-density state.

\emph{Temperature Induced Vitrification:} Figure~\ref{fig:glassTransition}D presents cooling trajectories at ambient pressure, initiated from 360~K and carried out at different cooling rates $q$. In analogy to Fig.~\ref{fig:glassTransition}A, slower cooling drives the system into progressively lower-energy glassy states, reflecting exploration of lower-lying minima in the potential-energy landscape. The inset shows isothermal aging at 195~K following the fastest cooling ramp. Over a timescale of $\sim$10~ns, the potential energy decreases by an amount comparable to the energy fluctuations observed in Fig.~\ref{fig:1}, confirming again, that these fluctuations reflect trapping in a glassy state rather than equilibrium liquid behavior.

Figure~\ref{fig:glassTransition}E compares the heat capacity $C_p = \partial H / \partial T$ extracted from the cooling simulations in panel D (see SI, Fig.~\ref{fig:enthalpy}) with the experimental heat capacity values. Filled circles denote experimental data for neat water, where the lowest-temperature values (olive symbols) were obtained from micron-sized droplets probed on ultrafast timescales to suppress crystallization \cite{pathak2021enhancement}. Experimentally, $C_p$ exhibits a maximum at 228.9~K, followed by a lower value at 228.5~K, indicative of a sharp anomaly. The simulated $C_p$ curves, color-coded according to cooling rate, reveal a systematic evolution: at the highest rate, $C_p$ is essentially constant until around 220~K where it then drops characteristic of a conventional glass transition, while slower cooling produces an additional peak that grows in amplitude and shifts to lower temperature, approaching the experimentally observed maximum. Since the simulated cooling rates remain approximately three orders of magnitude faster than those of the droplet experiments, even slower cooling would be expected to shift the peak further toward the experimental values.

For the slowest cooling rate, the peak of the $C_p$ occurs at $\sim$243 K. To demonstrate the correlation between this feature and the evolution of the glassy dynamics upon cooling, Figure~\ref{fig:glassTransition}F shows the temperature resolved mean-square displacement along the trajectory. At high temperatures, the water molecules are mobile, i.e., trMSD increases during cooling as they diffuse relative to their original position. At low temperatures, the molecules no longer diffuse, leading to a constant trMSD upon further cooling. The onset of vanishingly slow dynamics coincides with the $C_p$ maximum, indicating that vitrification cuts off the increase in $C_p$ and masks any putative thermodynamic divergence or extremum on the timescale imposed by the cooling rate. This result underlines the importance of distinguishing whether a feature like an extremum in a thermodynamic response function is a genuine liquid property or a kinetic effect, imposed by glassy dynamics.  

\subsection*{A Glass Transition Temperature in the Range $\sim$181--197~K}

The preceding analysis motivates an independent estimation of the glass transition temperature ($T_g$). Rather than characterizing the glass transition apparent within simulation time scales, we estimate the experimental $T_g$ by extrapolating well-equilibrated relaxation dynamics to the temperature $T_g$ at which the structural relaxation time reaches $\tau=100$~s, the conventional experimental definition of the glass transition. To this end we employ the Vogel–Fulcher–Tammann (VFT) form,
\begin{equation}
\tau = \tau_0 \exp\!\left[\frac{D}{T-T_0}\right],
\end{equation}
which captures the super-Arrhenius slowing down of supercooled liquids approaching the glass transition \cite{cavagna2009supercooled}. To probe the dynamics of the high- and low-density states observed in Fig.~\ref{fig:1}, we performed simulations at 1~bar and 5000~bar, corresponding to low- and high-density conditions, respectively, ensuring that the system remains outside the purported binodal across the full temperature range \cite{gartner2022liquid,szukalo2026energy}.

\begin{figure}[h!]
\centering
\begin{tikzpicture}

\node[anchor=north west] (img3) at (0,-1)
  {\includegraphics[width=0.7\textwidth]{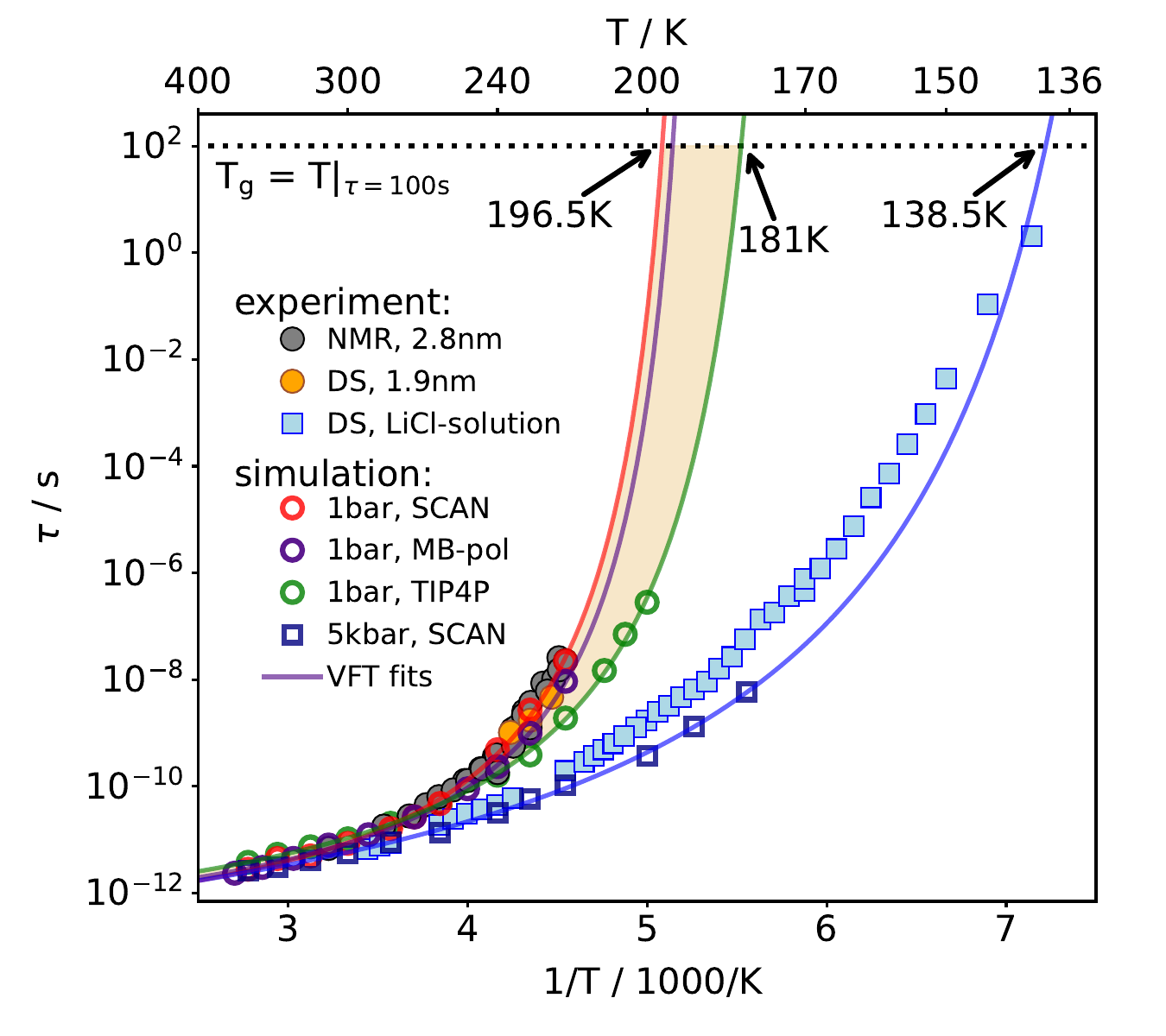}};
\node[anchor=north west] at (img3.north west) {\textbf{C}};

\node[anchor=south west] (img1) at ([xshift=1cm]img3.north west)
  {\includegraphics[width=0.29\textwidth]{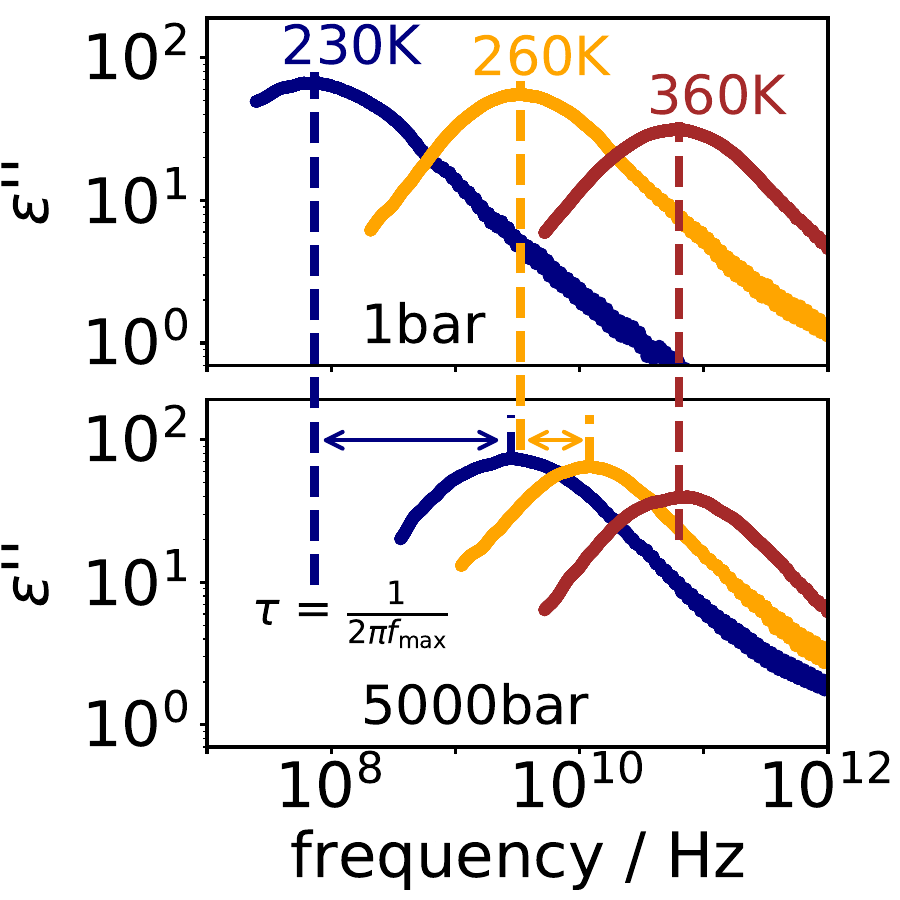}};
\node[anchor=north west] at (img1.north west) {\textbf{A}};

\node[anchor=north west] (img2) at (img1.north east)
  {\includegraphics[width=0.29\textwidth]{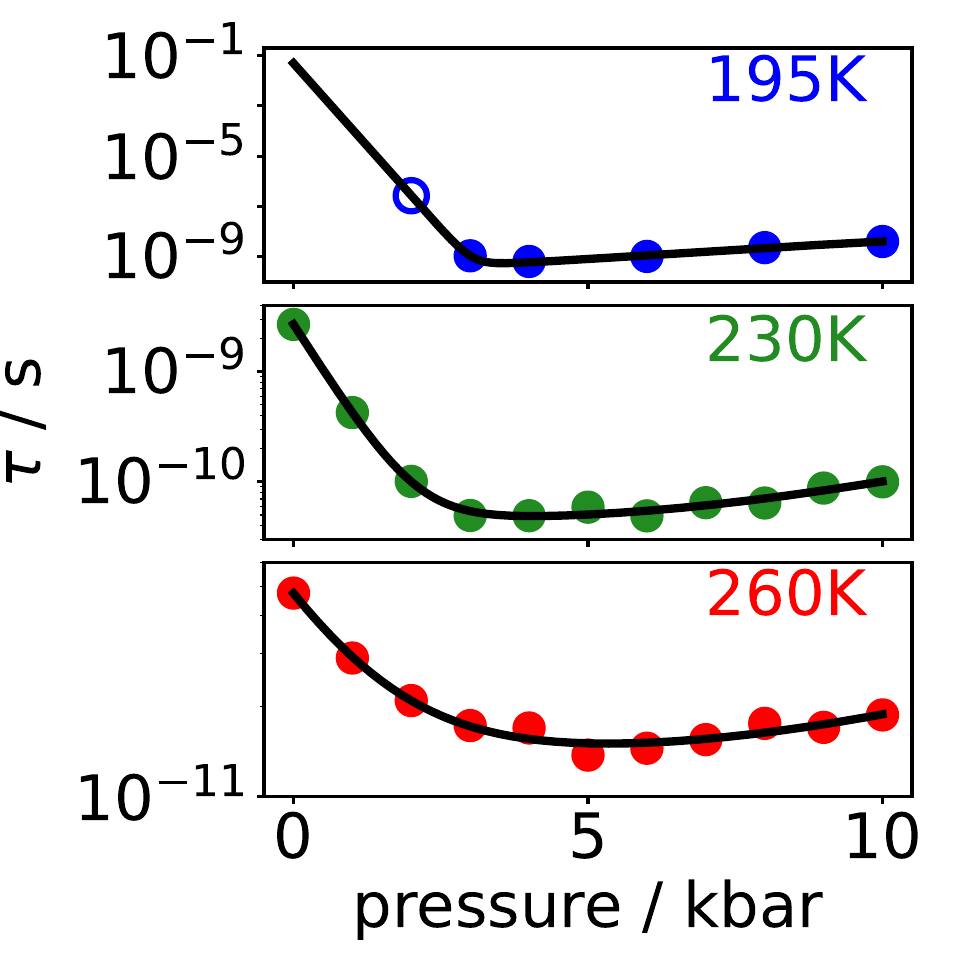}};
\node[anchor=north west] at (img2.north west) {\textbf{B}};

\end{tikzpicture}
\caption{\linespread{0.9}\selectfont \textbf{Pressure- and temperature-dependent structural relaxation}
(\textbf{A}) Dielectric-loss peak positions from simulations at 1~bar and 5000~bar separate upon cooling, revealing much faster slowing of structural relaxation at ambient pressure. 
(\textbf{B}) Corresponding relaxation times show a pronounced minimum between 3000 and 5000~bar, with rapid growth for lower pressures at low temperatures. Open symbol is an extrapolation and black line a guide to the eye, see SI, Fig.~\ref{fig:extrapol} for details.
(\textbf{C}) Comparison to experimental relaxation times from dielectric spectroscopy \cite{fischer2020supercooled} and NMR \cite{steinrucken2024complex} (vertically shifted by a factor of 2, see SI, Fig.~\ref{fig:total_self}) demonstrates close agreement with nanoconfined water at 1~bar. 
At 5000~bar, simulated relaxation times match those of a 14.8~mol\% LiCl solution \cite{lunkenheimer2025exploring} and yield a glass-transition temperature near 136~K. 
At 1~bar, relaxation slows much more steeply, leading to $T_g \approx 189\pm8$~K, depending on the model, placing vitrification in the same temperature range where a liquid--liquid transition has been proposed.
}
	\label{fig:dyn} 
\end{figure}

We extract the relaxation time $\tau$ by probing the structural relaxation of the liquid through its total electric dipole. This choice is motivated by two considerations. First, dipole fluctuations can be directly compared with dielectric spectroscopy, the most widely used experimental probe of supercooled-liquid dynamics. Dielectric spectroscopy spans an exceptional frequency range from terahertz to microhertz, thereby covering the entire crossover from liquid to glassy behavior. Importantly, dielectric spectroscopy measurements for water under nanoconfinement and in salt solutions---two strategies that suppress crystallization---are already available as a function of supercooling, enabling direct comparison with our simulations, as discussed below. Second, the dominant peak in water’s dielectric loss spectrum reflects collective orientational rearrangements of the hydrogen-bond network. This process occurs on longer timescales than translational diffusion over intermolecular distances\cite{arbe2016dielectric} and therefore sets the longest relaxation timescale required for full equilibration of the liquid. Accordingly, we show in the SI (Fig.~\ref{fig:arrh2}) that the viscosity is coupled to the rotational motions, while diffusion decouples from it at lower temperatures.

Figure~\ref{fig:dyn}A shows the temperature dependence of the peak position of the dielectric loss at 1~bar (top) and 5000~bar (bottom). While the dynamics at the two pressures are similar at high temperatures, they diverge progressively upon cooling, with relaxation at ambient pressure slowing much more rapidly than at high pressure, as highlighted by the arrows connecting corresponding peak positions. Figure~\ref{fig:dyn}B illustrates the resulting structural relaxation times, $\tau = (2\pi f_{\rm max})^{-1}$, as a function of pressure for three representative temperatures. At all temperatures, $\tau$ exhibits a pronounced minimum between 3000 and 5000~bar. Below this pressure range, relaxation slows increasingly steeply upon depressurization, an effect that becomes stronger at lower temperatures, precluding equilibration at low temperatures and pressures. Above the minimum, $\tau$ increases monotonically with pressure, as expected for a simple liquid. These trends mirror the pressure dependence of water viscosity observed experimentally\cite{singh2017pressure}.

In Figure~\ref{fig:dyn}C, relaxation times from simulations at 1~bar and 5000~bar are compared to experimental data obtained using dielectric spectroscopy (DS) \cite{fischer2020supercooled,lunkenheimer2025exploring} and nuclear magnetic resonance (NMR) \cite{steinrucken2024complex}. Experiments on nanoconfined water allow access to low temperatures but in some cases cease to probe true structural relaxation below a certain threshold due to partial crystallization \cite{cerveny2016confined,steinrucken2024complex}, finite-size effects \cite{fischer2020supercooled}, or the dominance of secondary relaxations \cite{melillo2024complexity}; these data are therefore omitted at temperatures lower than those accessed by the DNN simulations (220~K) for the sake of clarity, but are shown in Fig.~\ref{fig:diffusion}. In contrast, salt solutions effectively suppress crystallization and permit measurements down to the glass transition \cite{lunkenheimer2025exploring}. These solutions are known to resemble water under high pressure in many structural respects \cite{leberman1995effect}.

The simulated relaxation times at 5000~bar follow experimental results for a 14.8~mol\% LiCl solution. Fitting these data to the VFT form yields a glass-transition temperature of 138.5~K, remarkably close to the experimental value of 136~K for concentrated aqueous solutions \cite{lunkenheimer2025exploring}, showing that these solutions are a faithful analogue, both structurally\cite{leberman1995effect} and dynamically, for water under pressure.

By contrast, relaxation at 1~bar slows down far more steeply upon cooling and closely tracks the experimental data for nanoconfined water. The corresponding VFT fit yields $T_g = 196.5$~K for DNN@SCAN, placing vitrification near the temperature range where the putative liquid–liquid fluctuations are seen in Figure~\ref{fig:1}. This directly rationalizes the nearly flat dipole moment observed in the low-density regime in Figure~\ref{fig:1}, which arises from the orders-of-magnitude separation between the microsecond timescales accessible to the simulations and the equilibrium dynamics extrapolated to this temperature via the VFT fit. In the VFT description, relaxation times formally diverge at $T_0$, a temperature often associated with the Kauzmann temperature $T_K$ \cite{angell1997entropy}, where the extrapolated entropy of the liquid would fall below that of the crystal \cite{kauzmann1948nature}. For water, experimental estimates place $T_K$ between 180 and 190~K \cite{speedy1976isothermal}. From our fit we obtain $T_0 = 184.5$~K, in excellent agreement with these values. Similar analysis of trajectories from DNN@MB-pol yield a nearly identical $T_g = 194.5$~K and $T_0 = 183.2$~K, the latter in excellent agreement with the temperature where the excess entropy of this model vanishes \cite{szukalo2026energy} at similar density, see Fig.~\ref{fig:kauz} for details.
The TIP4P/2005 water model has a lower glass transition temperature of $181$~K, in accordance with the lower temperature (177~K) of the two-phase fluctuations \cite{debenedetti2020second}. Taking the average of the two limiting $T_g$ values, we arrive at a combined estimate of $189\pm8$~K.

\subsection*{Discussion and Conclusions}

The hypothesis of a liquid–liquid transition (LLT) has shaped the modern understanding of water’s anomalies. Recent experiments have advanced measurements deep into the no-man’s-land regime, and molecular simulations have provided independent evidence supporting critical behavior in supercooled water. By combining extensive machine-learning–accelerated first-principles simulations with direct comparison to multiple experimental measurements, we uncover a common dynamical origin underlying these interpretations: in the deeply supercooled regime near the proposed LLT, structural relaxation slows dramatically with small reductions in pressure or temperature. Systems commonly interpreted as liquid may therefore reside in a glassy state, biasing inferred thermodynamic properties. Mobilities in the low-density state for all three models in which two-state fluctuations have been observed—including DNN@MB-pol, currently the most accurate model for neutral bulk water—fall within the range reported in simulations for water in low-density amorphous (LDA) ice \cite{ghesquiere2015diffusion}. We propose that signatures associated with a liquid–liquid transition can also arise once the relaxation time of the system exceeds the observation timescale, such that equilibration becomes severely hindered and glassy behavior can resemble liquid–liquid coexistence.

\begin{figure}[h!]
    \centering
    \includegraphics[width=0.5\linewidth]{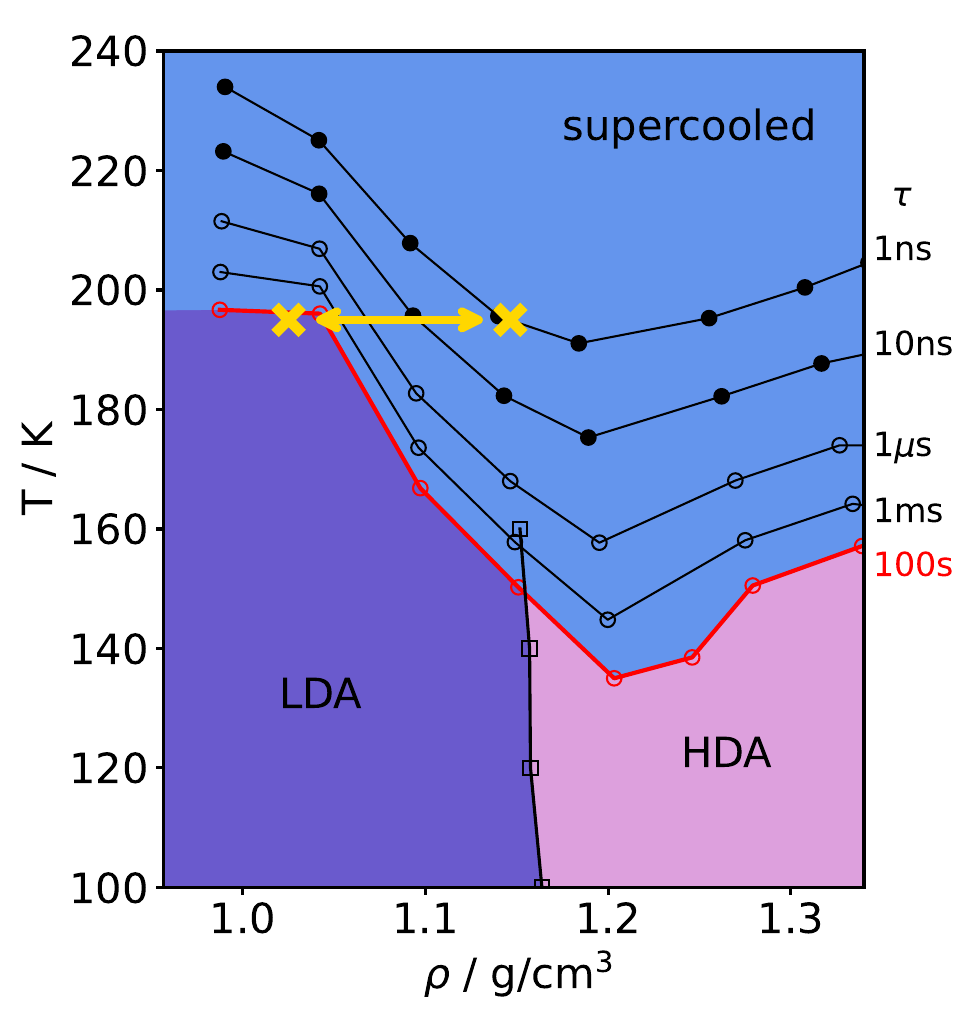}
    \caption{\textbf{"Phase diagram" of the DNN@SCAN model.} Circles connected by lines are isochrones, i.e., lines of equal structural relaxation time $\tau$, with values as indicated on the right side. Solid circles denote time scales accessible by the simulations, while open circles are extrapolations with the VFT equation, see Fig.~\ref{fig:dyn}. 
    Squares connected by lines is the first-order–like transition between the two amorphous phases, taken as the average of the pressure up- and down-ramps reported in \cite{szukalo2025computational}. 
    The two yellow crosses mark the densities of the low- and high-density states associated with the two-state fluctuations observed in Fig.~\ref{fig:1}.    
    }
    \label{fig:phase}
\end{figure}

To make this equilibration challenge more concrete, we construct a dynamical map of the supercooled phase diagram for the DNN@SCAN model by extracting isochrones, i.e., lines of constant structural relaxation time $\tau$ (Fig.~\ref{fig:phase}). These are obtained by fitting the Vogel–Fulcher–Tammann equation to $\tau(T)$ at different pressures and determining the temperatures at which $\tau$ reaches specified values. Filled circles denote data on timescales accessible by the simulations, while open circles indicate VFT extrapolations (see Fig.~\ref{fig:dyn}). The density along each isochrone is obtained by inter- or extrapolation of $\rho(T,P)$ using a polynomial fit (see SI Sec.~\ref{sec:SI-density} for details). For reference, we also include the densities at which the transition between low- and high-density amorphous ice has been reported \cite{szukalo2025computational}, shown as black open squares. These values correspond to the average transition density obtained from both compression and decompression ramps. The two yellow crosses, connected by an arrow, mark the densities of the low- and high-density states associated with the two-state fluctuations observed in Fig.~\ref{fig:1}.

Several key features emerge from this phase diagram. At densities above approximately $1.2~\mathrm{g/cm^3}$, water exhibits conventional liquid behavior: densification results in isochrones with a positive slope, indicating slower dynamics at higher density and constant temperature. In contrast, below $\sim 1.2~\mathrm{g/cm^3}$, anomalous behavior dominates, characterized by steep isochrones with negative slope. Most notably, the two-state fluctuations reported in Fig.~\ref{fig:1} occur in a regime where the isochrones are extremely closely spaced in density. In this narrow region, a density difference of only $\sim 0.12~\mathrm{g/cm^3}$ gives rise to an approximately 11-order-of-magnitude change in relaxation timescales, placing the low-density state within the low-density amorphous ice (LDA) regime. This behavior is less apparent in a pressure--temperature representation, as the pressure remains constant during the two-state fluctuations, thereby obscuring the discontinuous shift along the isochrones, rendering the assessment of LLCP equilibration in the $T$--$P$ plane nontrivial \cite{szukalo2025computational}.

As noted above, such fluctuations are observed only in small systems; larger-scale simulations \cite{malosso2024evidence, wang2026direct}, as well as simulations employing alternative barostat settings (see SI Fig.~\ref{fig:barostat}), do not exhibit this behavior. The precise origin of these two-state fluctuations offers several possibilities. Assuming one can invoke thermodynamic equilibrium, then these density fluctuations can be rationalized with the existence of the LLCP. On the other hand, if the breakdown of ergodicity at these conditions is confirmed, then it likely reflects a combination of factors, including the large disparity in relaxation times between the two states, barostat effects, and finite-size contributions. Moreover, how these factors influence free-energy-barrier calculations between the two states remains an open question.

The pressure dependence of water’s dynamics under supercooling thus reveals a complex landscape of structural relaxation. At higher pressures, the dynamics resemble those of concentrated aqueous solutions, with a glass-transition temperature near 136 K, consistent with previous reports \cite{lunkenheimer2025exploring}. At ambient pressure, however, relaxation slows much more dramatically. This stronger slowdown at low pressures can be rationalized by considering the distinct structural evolution upon cooling. At high pressures, the dynamics are governed primarily by conventional glassy slowdown. In contrast, at low pressures, cooling is accompanied by a marked structural transformation toward a more ordered, tetrahedral hydrogen-bond network. This enhanced structural ordering imposes additional constraints on molecular rearrangements, leading to a further reduction in dynamics through restricted hydrogen-bond exchange \cite{stirnemann2012communication}, a behavior not observed in simple glass-forming liquids. As a consequence, the extrapolated glass-transition temperature at ambient pressure is surprisingly high, $T|_{\tau=100\mathrm{s}} = 189\pm8$ K in the DNN@SCAN and DNN@MB-pol models as well as TIP4P/2005. This value agrees with recent studies of confined water \cite{melillo2024complexity}, rapidly cooled water \cite{kruger2023electron}, and salt mixtures that minimally perturb water’s structure \cite{zhao2016apparent,woutersen2018liquid}. 

Beyond rotational and structural relaxation, translational dynamics display closely related anomalies. Experimental diffusion coefficients, inferred from crystal growth rates, exhibit a clear kink in their temperature dependence around 189 K \cite{xu2016growth}. This feature may be interpreted as a transition from crystallization advancing via structural ($\alpha$) relaxation at higher temperatures to proceeding via secondary ($\beta$) relaxation below $T_g$ (see SI, Fig.~\ref{fig:diffusion}) \cite{swenson2018possible}. Because not only crystal growth rates but also direct diffusivity measurements \cite{kimmel2025translational} and dielectric measurements \cite{amann2013water} share the same temperature dependence at the lowest temperatures, an alternative interpretation is that the kink originates from non-equilibrium relaxation, as suggested by calculations based on facilitation theory \cite{limmer2014length}. In either case, these experiments are consistent with probing the onset of out-of-equilibrium dynamics in the vicinity of 189 K.

Together with our simulation results, these experimental observations suggest that signatures of an abrupt, first-order–like transition between high- and low-density states can emerge when at least one of these states falls out of equilibrium. Under equilibrium conditions, by contrast, the density varies smoothly with pressure and temperature, as the system remains able to adapt to the imposed state point. This interpretation is also consistent with recent coexistence simulations, in which a low-mobility, low-density phase of water is brought into contact with a high-mobility, high-density phase \cite{wang2026direct}. In these simulations, the interface persists for tens of nanoseconds—timescales comparable to those associated with aging toward lower-energy states in deeply supercooled systems (insets in panel A and D of Fig.~\ref{fig:glassTransition}). Similar front-like melting dynamics have been reported in glassy systems \cite{flenner2019front}, and model glass formers can exhibit first-order–like coexistence between ordered glassy and disordered liquid phases \cite{campo2020dynamical}. Our perspective is consistent with earlier suggestions that a relatively
high glass-transition temperature in the vicinity of the proposed LLT
would introduce a dynamical confound that any thermodynamic
interpretation of the observed signatures must address
\cite{sciortino1996supercooled}, as well as with models in which a
liquid--liquid critical point lies below the Kauzmann temperature
\cite{sciortino2003physics}, leaving the possible influence of such a
thermodynamically inaccessible, or \emph{virtual}, critical point an open question.

Viewed in this light, a crossover from a high-density liquid to a glassy low-density state offers a complementary rationalization for the inferred two-liquid behavior. Our findings therefore suggest that singularity-free scenarios for water’s anomalies \cite{sastry1996singularity,caupin2021minimal}, in which anomalous thermodynamic behavior arises from the continuous evolution of local structure rather than from a hidden critical point, merit renewed consideration. It is notable that medium-density amorphous ice (MDA) has recently been produced experimentally \cite{rosu2023medium}, indicating that the variety of structures at low temperatures may be richer than a simple two-state picture would suggest. In this context, it has also been shown that molecular reorientational motions underpin the mechanisms behind glass transitions in amorphous ices\cite{salzmann2016}. Our analysis demonstrating the importance of collective reorientational fluctuations in ergodicity breaking under supercooling is consistent with this observation. Moreover, several models reproduce many of water’s anomalies in the absence of a liquid–liquid critical point, including the coarse-grained mW model \cite{holten2013nature}. Maxima in thermodynamic response functions, resembling a Widom line, are also expected
within singularity-free scenarios for substances such as water, which
expand upon cooling \cite{sastry1996singularity}. Distinguishing these maxima at equilibrium from signatures arising from non-equilibrium effects or from an LLT remains an open challenge. More broadly, the phenomenology traditionally attributed to a liquid–liquid transition can also emerge from the rapid onset of a non-equilibrium, glassy state; from this perspective, the longstanding debate over supercooled water may reflect not only a critical point, but also the boundary between liquid and glassy states, which gives rise to phenomena that closely resemble a liquid–liquid transition.



%

\bibliography{bib} 
\bibliographystyle{sciencemag}



\section*{Acknowledgments}

We are indebted to Elisa Steinrücken for providing the NMR data from Ref.~\cite{steinrucken2024complex}, and to Francesco Paesani and Francesco Sciortino for providing the trajectories from Ref.~\cite{sciortino2025constraints} and Ref.~\cite{debenedetti2020second}, respectively.
We are grateful to Stefano Baroni, Pablo Debenedetti, Greg Kimmel, Francesco Sciortino, Pablo Piaggi and Robin Horstmann for helpful discussions. F.P. acknowledges support by the European Commission through the MAX Centre of Excellence for supercomputing applications (Grant No. 101093374), by
the Italian MUR, through the PRIN Project ARES (Grant
No. 2022W2BPCK), and by the Italian National Centre
for HPC, Big Data, and Quantum Computing (Grant No.
CN00000013), funded through the Next generation EU
initiative. A.H. acknowledges funding from the European Research Council (ERC) under the European Union’s Horizon 2020 research and innovation programme (grant agreement No. 101043272 – HyBOP). The views and opinions expressed are those of the authors only and do not necessarily reflect those of the European Union or the European Research Council Executive Agency. Neither the European Union nor the granting authority can be held responsible for them.

\comments{

\paragraph*{Funding:}
List the grants, fellowships etc. that funded the research;
use initials to specify which author(s) were supported by each source.
Include grant numbers when appropriate or required by the funding agency.
For example: F.~A. was funded by the Generous Science Agency grant~2372.
\paragraph*{Author contributions:}
List each author’s contributions to the paper.
Use initials to abbreviate author names.
\paragraph*{Competing interests:}
Disclose any potential conflicts of interest for all authors, such as patent applications,
additional affiliations, consultancies, financial relationships etc.
See the journal editorial policies web page for types of competing interest that must be declared.
If there are no competing interests, state:
``There are no competing interests to declare.''
\paragraph*{Data and materials availability:}
Specify where the data, software, physical samples, simulation outputs or other materials
underlying the paper are archived.
They must be publicly accessible when the paper is published (without embargo) and enable
readers to reproduce all the results in the paper.
Contact the editor if you’re unsure what needs to be shared.

Our preference is for digital material to be deposited in a suitable non-profit online data or
software repository that issues the material with a DOI.
Alternatively, an institutional repository, subject-based archive, commercial repository etc.
is acceptable, as are (short) supplementary tables or a machine-readable supplementary data file.
‘Available on request’ or personal web pages are not allowed.

Cite the relevant DOI \cite{dataset}, URL \cite{example_url} or reference \cite{example2}
in this statement.
These \textit{do not} count towards the reference limit if they are only cited in the acknowledgements.
Be specific and state a unique identifier -- such as an accession number, software version number
or observation ID -- so readers can easily retrieve the exact material used.

Declare any restrictions on sharing or re-use -- such as a Materials Transfer Agreement (MTA) or
legal restrictions -- which must be approved by your editor.
Unreasonable restrictions will preclude publication.
Consult the journal's editorial policies web page for more details.
}




\newpage


\renewcommand{\thefigure}{S\arabic{figure}}
\renewcommand{\thetable}{S\arabic{table}}
\renewcommand{\theequation}{S\arabic{equation}}
\renewcommand{\thepage}{S\arabic{page}}
\setcounter{figure}{0}
\setcounter{table}{0}
\setcounter{equation}{0}
\setcounter{page}{1} 


\begin{center}
\section*{Supplementary Information for\\ \scititle}

Florian Pabst$^{\ast}$,
Ali Hassanali$^{\ast}$\\ 
\small$^\ast$Corresponding author. Email: fpabst@sissa.it; ahassana@ictp.it\\

\end{center}





\section{Materials and Methods}
Three different water models were used for the simulations in this study, for which two-state fluctuations have previously been reported and attributed to a liquid--liquid transition: the classical TIP4P/2005 model \cite{abascal2005general}, and two deep neural network potentials trained either on DFT data (DNN@SCAN \cite{zhang2021phase}) or on MB-pol (DNN@MB-pol \cite{bore2023realistic}). 

All temperatures reported in this work for simulations employing the DNN models are shifted by the difference in melting temperature between experiment and the respective model ($-40$~K for DNN@SCAN \cite{zhang2021phase} and $+10$~K for DNN@MB-pol \cite{bore2023realistic}) in order to facilitate comparison with experiment. It is well known that, after applying such a shift, the dynamics obtained from DNNs and experiment are in good agreement \cite{pabst2025glassy,pabst2025salt}.

Simulations with the DNNs are performed using LAMMPS \cite{thompson2022lammps}, interfaced with the DeepMD-kit \cite{zeng2023deepmd}, on systems of 512 water molecules. For all simulations from which correlation times $\tau$ are deduced, great care is taken to equilibrate the system prior to production runs, as follows. First, NPT simulations are performed, starting from the final configuration of the preceding higher-temperature run, for at least $10\tau$ with a time step of 0.5~fs. The density obtained at the end of this simulation is then used for a subsequent NVT run, again with a 0.5~fs time step. After approximately $10\tau$, eight independent configurations, separated by at least $1\tau$, are extracted and used as starting points for eight production runs in the NVE ensemble, with a time step of 0.25~fs and a duration of approximately $10\tau$. The lengths of all production runs are given in table~\ref{table:simLength}.

Simulations with the TIP4P/2005 model are performed in GROMACS on systems of 1000 water molecules, using the same settings as in the study reporting two-state fluctuations \cite{debenedetti2020second}. The same NPT--NVT--NVE protocol described above is employed, but with a time step four times larger. At the lowest temperatures (220~K and below), the temperature could not be maintained within acceptable limits during the very long NVE runs. Therefore, in these cases, production runs are performed in the NVT ensemble using the v-rescale thermostat \cite{bussi2007canonical}.

Dielectric spectra (imaginary part of the dielectric function) are calculated from the total electric dipole moment of the simulation box, obtained using a second neural network trained to predict molecular dipole moments \cite{malosso2024evidence}, via the equation
\begin{equation}
    \varepsilon''(\omega) = \frac{\omega}{3V\varepsilon_0k_BT}\int \text{d}t \exp(-i\omega t) \left< \mathbf{M}(t)\mathbf{M}(0) \right>,
    \label{eq:eps2}
\end{equation}
where $V$ is the box volume, $\mathbf{M}$ is the total dipole moment of the system (i.e., the sum over all molecular dipole moments), $T$ is the simulation temperature, and $\epsilon_0$ and $k_B$ are the vacuum permittivity and Boltzmann constant, respectively. This method has been shown to yield excellent agreement with the experimental dielectric spectrum of ambient water; further details can be found in Ref.~\cite{pabst2025salt}.

{\tiny
\begin{table}
    \centering
    \resizebox{\textwidth}{!}{
    \begin{tabular}{c|ccccccccccccccccccc}
        T / K & 370 & 360 & 350 & 340 & 330 & 320 & 310 & 300 & 290 & 280 & 270 & 260 & 250 & 240 & 230 & 220 & 210 & 205 & 200 \\ \hline
        SCAN & -& 0.3 & - & 0.4 & - &  0.6 & - & 1.1 & - & 2.2 & 4 & 7.8 & 12.8 & 32 & 102.4 & 140.8 & - & - & - \\
        MB-pol & 0.5 & - & 0.5 & - & 0.75 & - & 1.25 & - & 2.5 & - & 4 & - & 4 & 25.6 & 51.2 & 76.8 &  - & - & - \\
        TIP4P & - & 0.5 & - & 0.5 & - &  0.5 & - &  1 & - & 2 & - & 4 & - & 8 & 16 & 32 & 256 & 2560 & 5120  \\
    \end{tabular}}
    \caption{Simulation time of the production runs in nanoseconds for the DNN@SCAN, DNN@MB-pol and TIP4P/2005. }
    \label{table:simLength}
\end{table}
}


\section{Supplementary Information for Fig.~1}

\subsection{Trajectories from the literature}
\begin{figure}[htb!]
\includegraphics[width=0.99\textwidth]{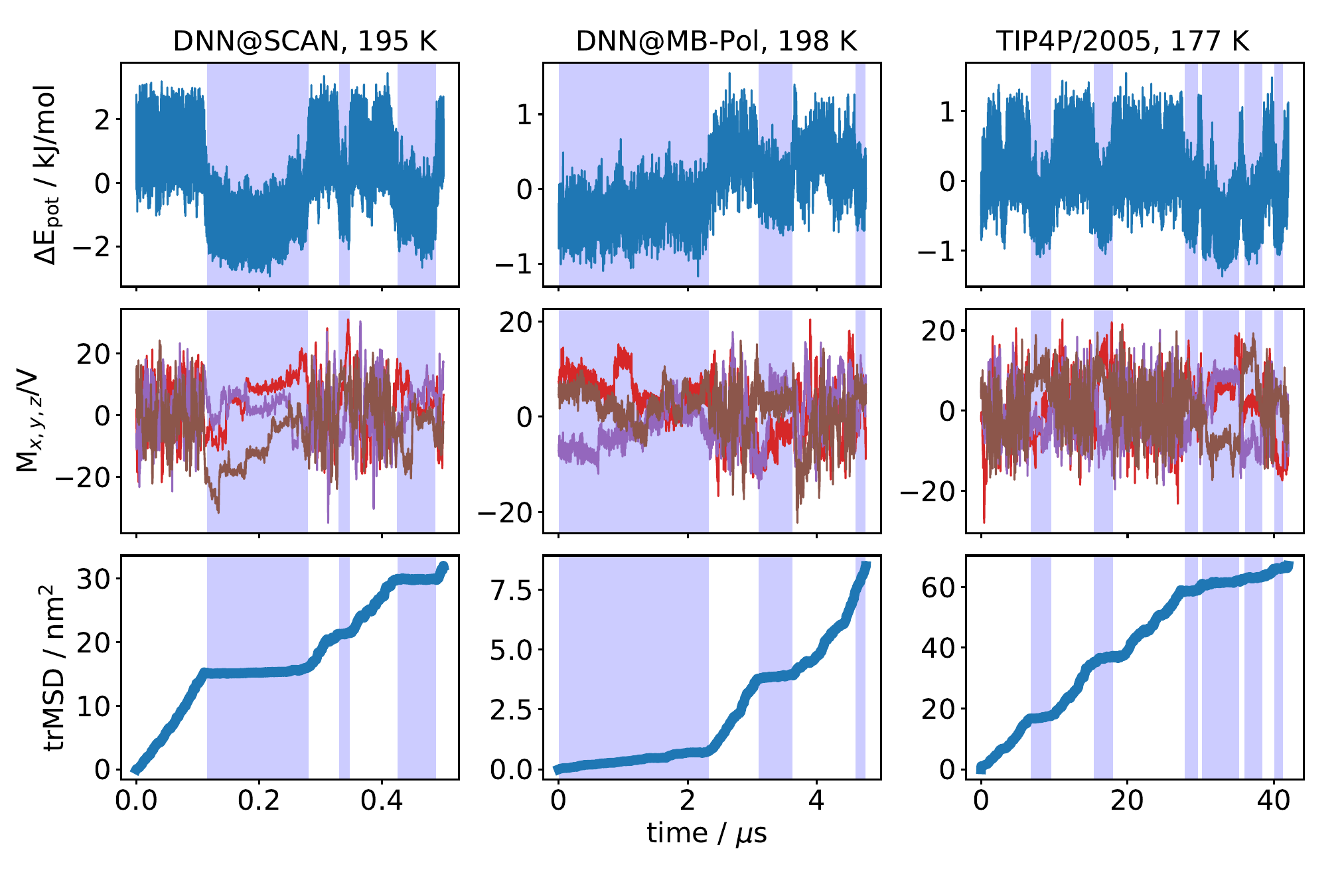}
\caption{\linespread{0.9}\selectfont \textbf{Two-state fluctuations from the literature.}
Trajectories from the literature \cite{gartner2022liquid,sciortino2025constraints,debenedetti2020second} obtained with two neural-network water models (DNN@SCAN, DNN@MB-pol) and the classical TIP4P/2005 potential show alternating intervals of high and low potential energy (in kJ/mol per molecule, upper row). Rotational and translational dynamics are quantified via the total electric dipole components in three cartesian directions (middle row) and the trMSD (lower row, see eq.~\ref{eq:trMSD}), respectively. Regimes characterized by low energy values exhibit notably reduced dynamics.   
Temperatures indicated for the DNN models are corrected for their melting point mismatch with respect to experiment.}
\label{fig:MSD2}
\centering
\end{figure}
Figure~\ref{fig:MSD2} presents three representative trajectories from the literature obtained with two state-of-the-art deep neural-network (DNN) potentials, DNN@SCAN\cite{gartner2022liquid} and DNN@MB-pol\cite{sciortino2025constraints} and the classical TIP4P/2005 model \cite{debenedetti2020second}. In analogy to Fig~\ref{fig:1}, the transformation between the HDL and LDL phases are reflected in fluctuations observed in the potential energy of the system illustrated in the top row for the three models on the microsecond timescale. The rotational and translational dynamics are shown in the middle and lower row, respectively. Like in the main text, rotational motions are tracked via the total electric dipole moment of the simulation box in three cartesian directions and translational motions via the time resolved mean square displacement (see eq.~\ref{eq:trMSD}). In accordance with Fig.~\ref{fig:1}, regimes of low potential energy exhibit notably reduced dynamics compared to regimes of high potential energy. This is evident from both the dipole and the trMSD traces: While the dipole components fluctuate rapidly around zero in the high energy (high density) state, long lived polarization develops in the low energy (low density) states. Likewise, the trMSD increases linearly in the high energy states, signaling molecular diffusion, while it stays essentially flat in the low energy regimes, as a sign of notably reduced diffusion.

\begin{figure}[h!]
    \centering
    \includegraphics[width=0.7\linewidth]{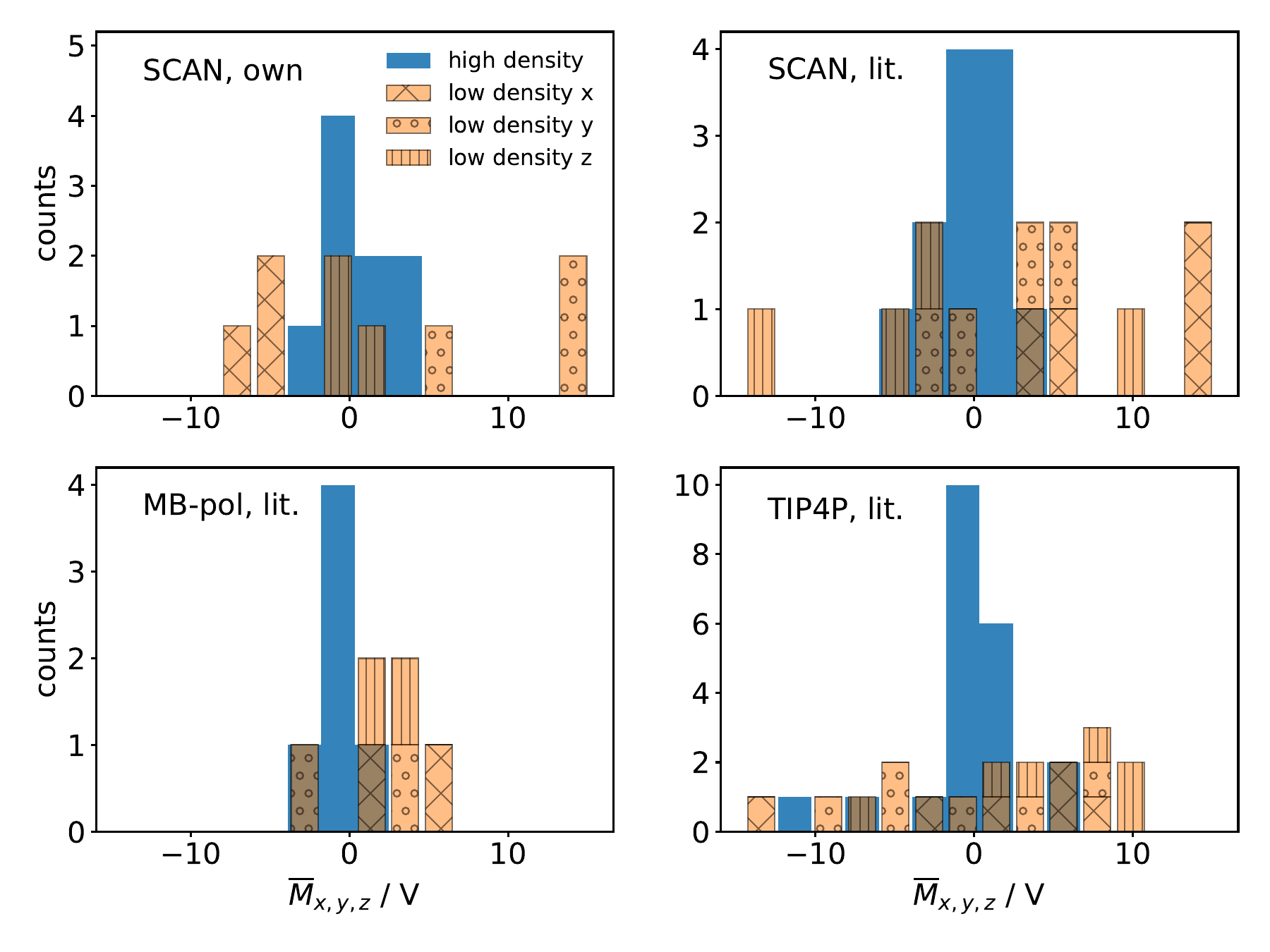}
    \caption{Mean dipole moments (in units of Debye/nm$^3$) for high and low density segments of the trajectories shown in Fig.~\ref{fig:1} and Fig.~\ref{fig:MSD2}. A value of zero is expected for a liquid in equilibrium, while the distributed values in the low density state are a sign of broken ergodicity.}
    \label{fig:pol}
\end{figure}

To appreciate the extend of broken ergodicity in the low-energy state, it is instructive to calculate the mean dipole value for each shaded segment in Fig.~\ref{fig:MSD2} in three cartesian directions: For an equilibrated liquid, these mean components must be equal to zero, since any persistent nonzero average reflects incomplete sampling of configuration space and thus a loss of ergodicity. For all three trajectories from the literature shown in Fig.~\ref{fig:MSD2} as well as for our own trajectory in Fig.~\ref{fig:1} we calculate the mean mean dipole for each segment in three cartesian coordinates and show the result in Fig.~\ref{fig:pol}. The borders between segments are identified by a change point detection method implemented in the \textit{ruptures} python package \cite{truong2020selective} monitoring changes in the mean and variance of the density. For the high density states, most of the dipole moments indeed average to zero, indicating liquid behavior, exceptions are seen when the segments are rather short. In stark contrast, the average dipole moments of the low-density segments are mostly different from zero, with values distributed from -14 to 14 in units of Debye per nm$^3$. This observation signals that these low-density segments can not be regarded as equilibrated liquid samples, but have to be considered non-equilibrium glassy states where dynamics are slow and ergodicity is broken. Thus, the fluctuations between the two states can not be regarded as an equilibrium phase transition and great care has to be taken when trying to infer for instance a free energy landscape from these trajectories. 

\subsection{Influence of the barostat on two-state fluctuations}

\begin{figure}[h!]
     \centering
     \includegraphics[width=0.85\linewidth]{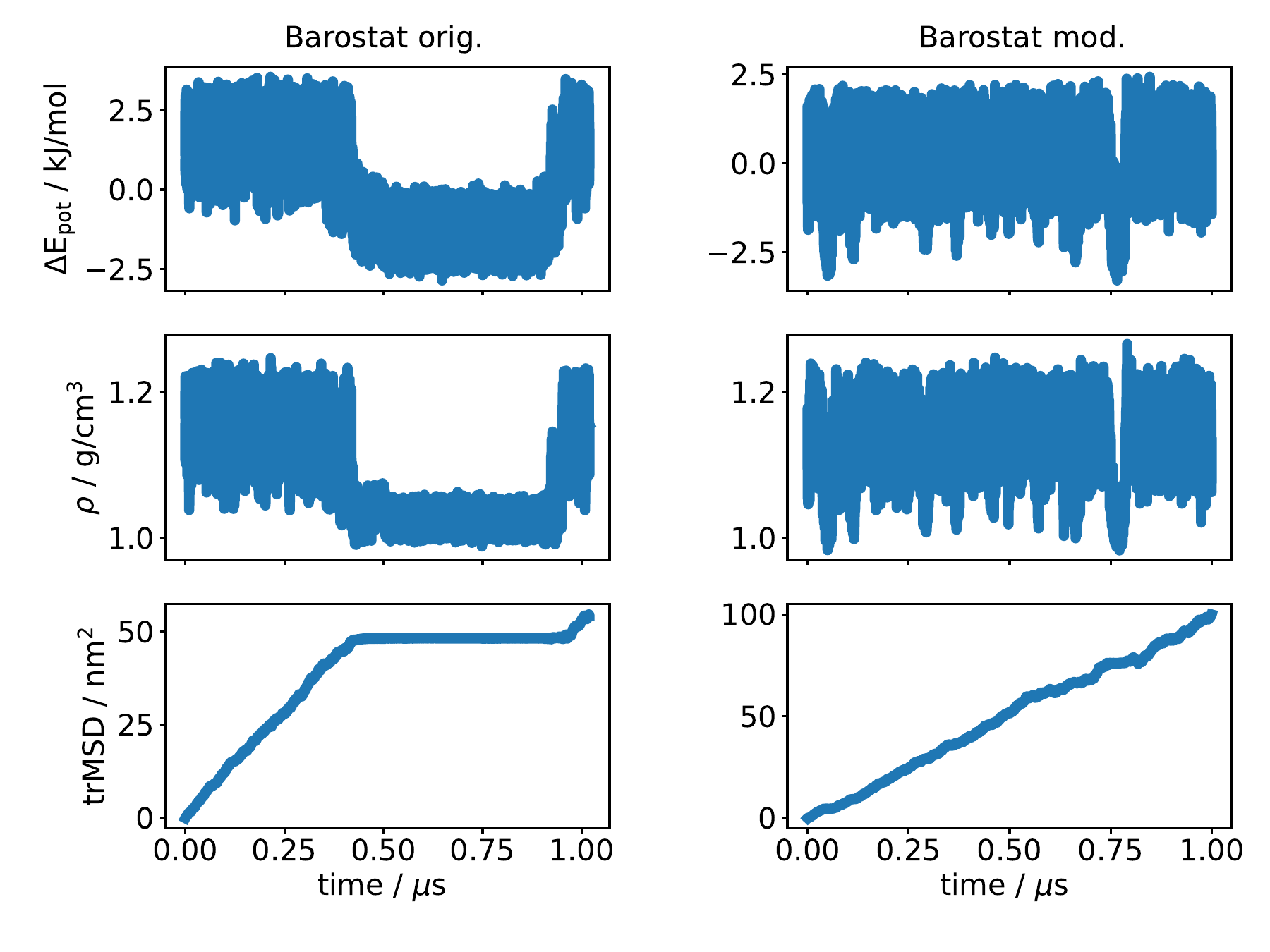}
     \caption{\textbf{Influence of barostat settings on two-state fluctuations.} Simulations performed with the DNN@SCAN model and 192 water molecules. Left: All setting according to the original work \cite{gartner2022liquid}, where two-state fluctuations were observed (see Fig.~\ref{fig:MSD2}). Right: Damping times of barostat and thermostat increased by a factor of 10, all other settings identical.}
     \label{fig:barostat}
 \end{figure} 

In the main text, we showed that the two-state fluctuations, which were previously assigned to a liquid--liquid transition, instead reflect intermittent trapping in a low-energy glassy state that is not accessible in equilibrium at the corresponding state point. 
Since large pressure fluctuations in the small simulation box (192 molecules) are presumed to facilitate trapping in an energy state corresponding to lower pressures, we investigated how different barostat settings influence the observed two-state fluctuation behavior. To this end, we compare the simulation presented in Fig.~\ref{fig:1}, where we used identical settings and procedures (3200~bar, initialized in the high-density regime) as in the literature \cite{gartner2022liquid} to detect the purported liquid--liquid fluctuations, with a second simulation in which all parameters were kept identical, except that the damping constants of both the barostat and thermostat were increased by a factor of 10 (i.e., $\tau_{\rm thermo}$ from 50~fs to 500~fs and $\tau_{\rm baro}$ from 500~fs to 5~ps). 
Both simulations were run for approx. 1~$\mu$s, i.e., twice as long as in the original work. The potential energy, density, and time-resolved mean square displacement are shown in Fig.~\ref{fig:barostat}, with the original settings on the left-hand side and the modified settings on the right-hand side. While the simulation performed with the original settings exhibits trapping in a low-energy, low-density, low-mobility state for a significant fraction of the trajectory (here, $\sim 0.5~\mu$s), as described in the main text, the behavior is markedly different for the modified settings: In this case, no long-lived trapping in the low-energy state is observed; instead, only very short excursions occur, with the longest appearing at around 0.75~$\mu$s and lasting for approximately $\sim 50$~ns. 
These results demonstrate that the two-state fluctuations are highly sensitive to barostat parameters and may therefore not be indicative of an underlying equilibrium liquid--liquid transition, but instead from non-equilibrium trapping effects.


\section{Supplementary Information for Fig.~2}
\subsection{Dynamic slowing-down during depressurizing/cooling}
\begin{figure}[h!]
    \centering
    \includegraphics[width=0.45\linewidth]{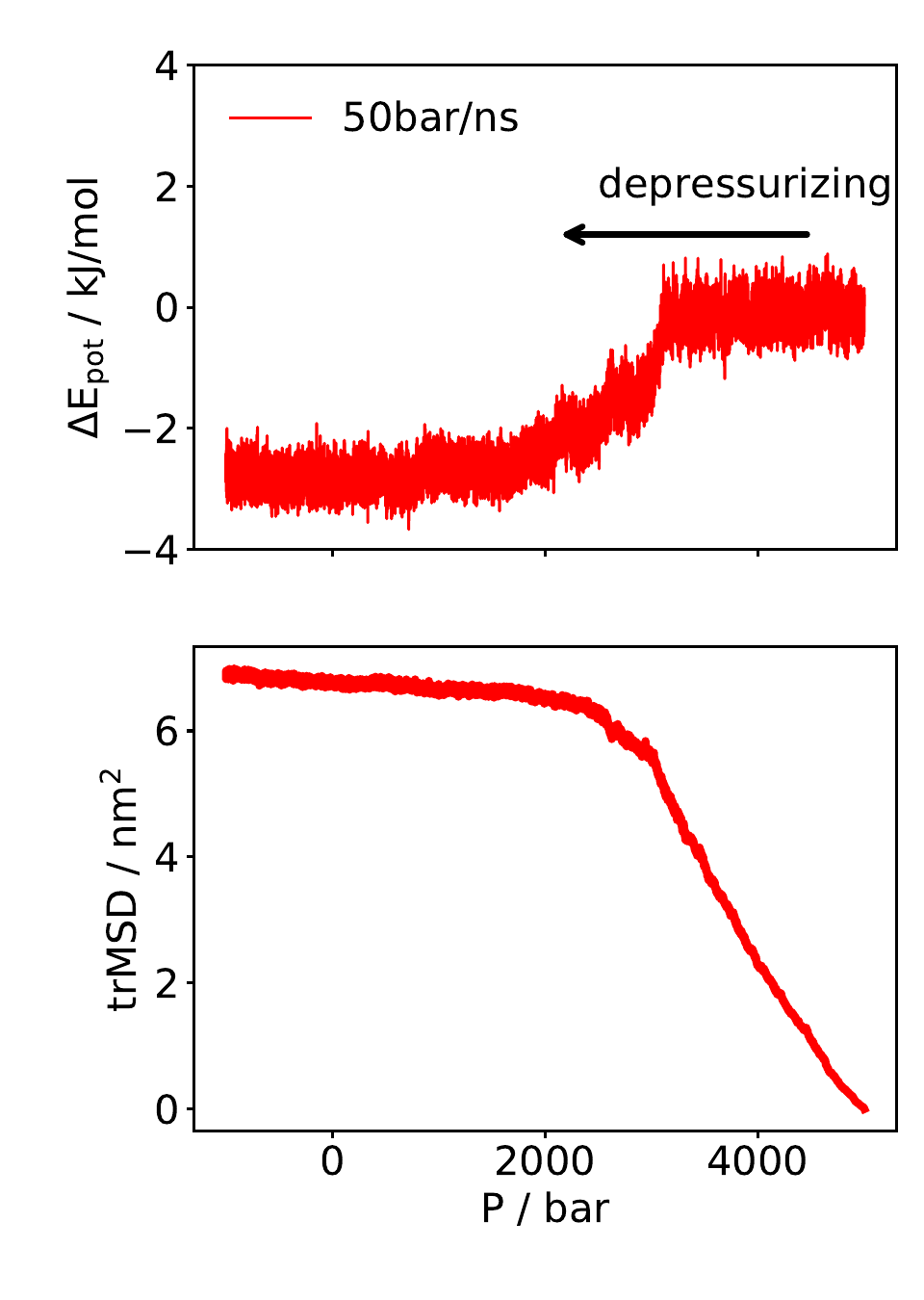}
    \includegraphics[width=0.45\linewidth]{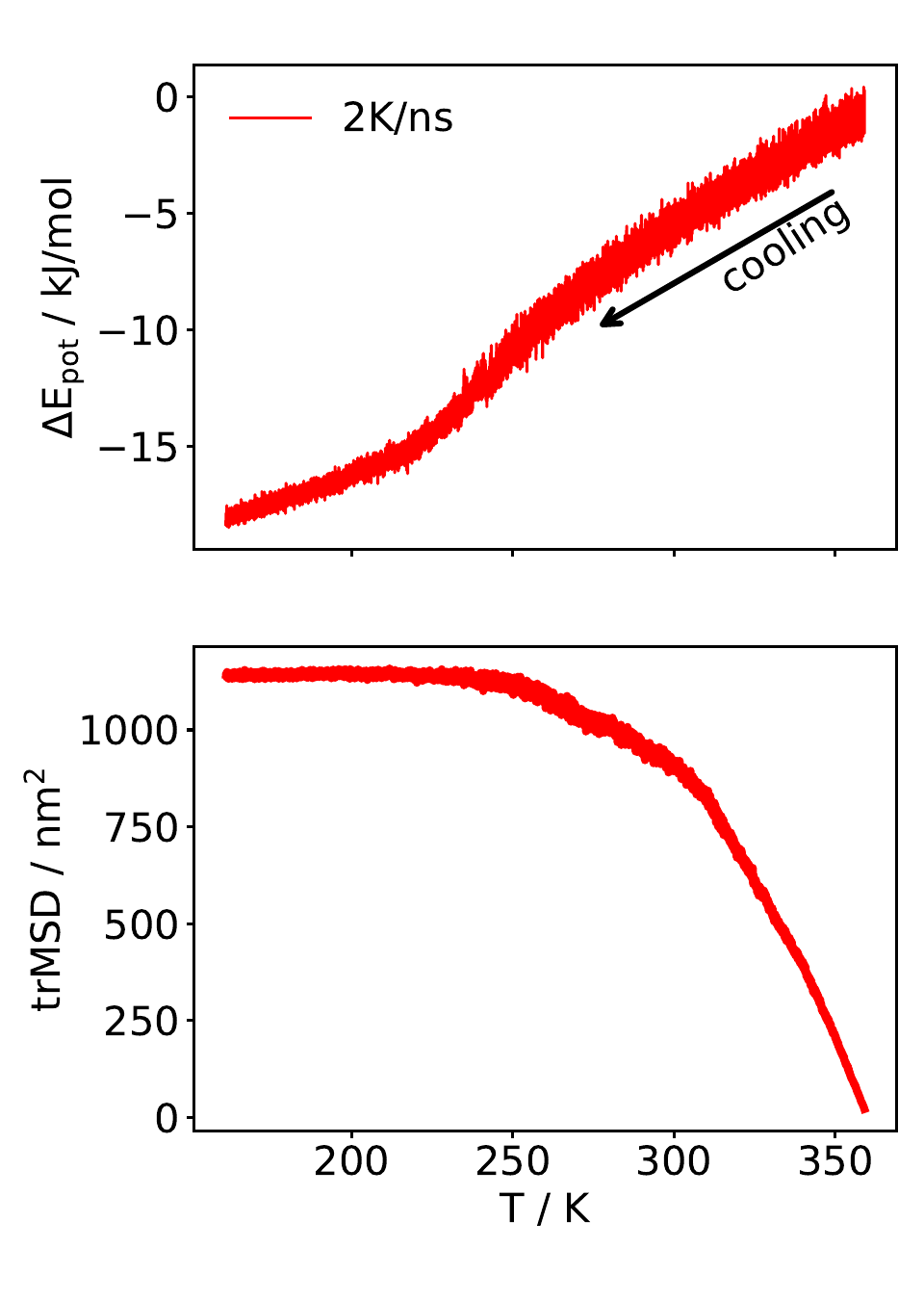}
    \caption{\textbf{Dynamic slowing-down during pressure and temperature ramps.} For both pressure and temperature ramps slowing dynamics (as inferred from a constant trMSD) occur at the same time as the system transitions to a low-energy state.}
    \label{fig:dynArrest}
\end{figure}

In Fig.~\ref{fig:dynArrest}, we reproduce the evolution of the potential energy during the depressurization run at 50~bar/ns and the cooling run at 2~K/ns shown in Fig.~\ref{fig:glassTransition}, and compare it to the instantaneous mean square displacement. All curves are to be read from right to left. At high pressures (temperatures), the molecules are mobile, as evidenced by the increase in the time-resolved mean square displacement (trMSD) upon depressurization (cooling). At the point where a distinct transition to a lower-energy state occurs, the trMSD levels off and becomes essentially constant at low pressures (temperatures), signaling vanishingly slow dynamics in a glassy state.

\subsection{Heat capacity calculation}

\begin{figure}[h!]
    \centering
    \includegraphics[width=0.5\linewidth]{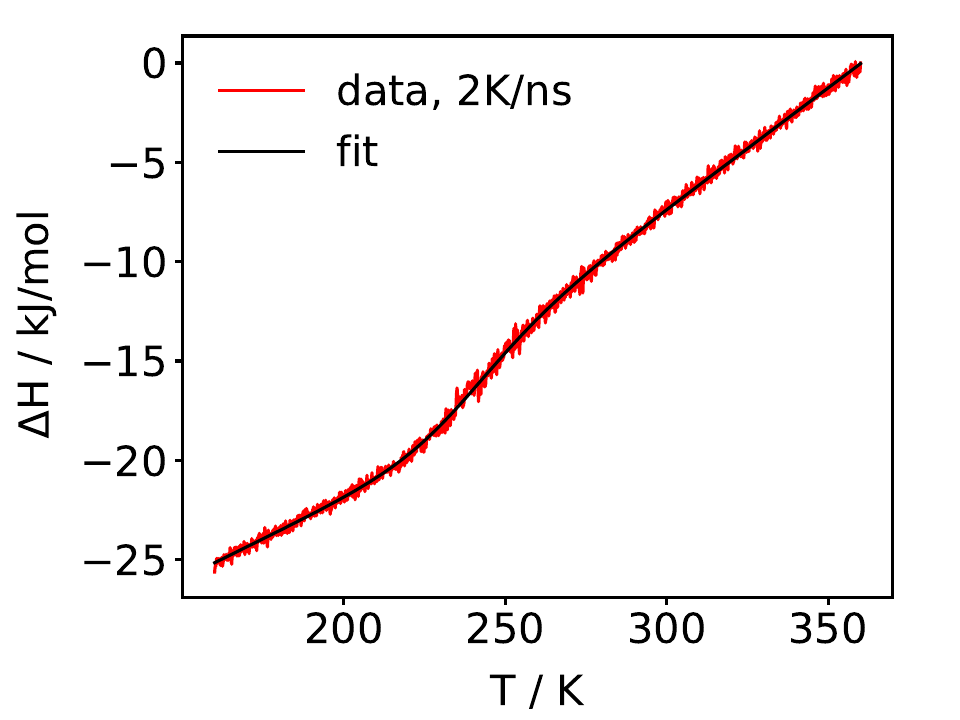}
    \caption{\textbf{Enthalpy fitting.} Fit of the enthalpy data during cooling with equation~\ref{eq:enthalpy} to yield a smooth representation for calculating the derivative. }
    \label{fig:enthalpy}
\end{figure}

In order to obtain the smooth heat capacity ($C_p$) curve shown in Fig.~\ref{fig:glassTransition}E from the noisy enthalpy $H$ data during cooling via $C_p = \partial H/\partial T$, we fit the enthalpy data with an empirical equation of the form

\begin{equation}
\Delta H = (a_1+a_2T) +  \frac{(a_3+a_4T)-(a_1+a_2T)}{1+\exp(-(T-T_r)/s)},   
\label{eq:enthalpy}
\end{equation}
which we found to describes the data very well, where $a_1,a_2,a_3,a_4,s,T_r$ are fitting parameters. The quality of the fit is shown in Fig.~\ref{fig:enthalpy}. We use this fit for the noise-free calculation of the derivative.

\section{Supplementary Information for Fig.~3}

\subsection{Panel A/C: Dielectric vs. NMR correlation times} \label{sec:NMR}
\begin{figure}[h!]
    \centering
    \includegraphics[width=0.5\linewidth]{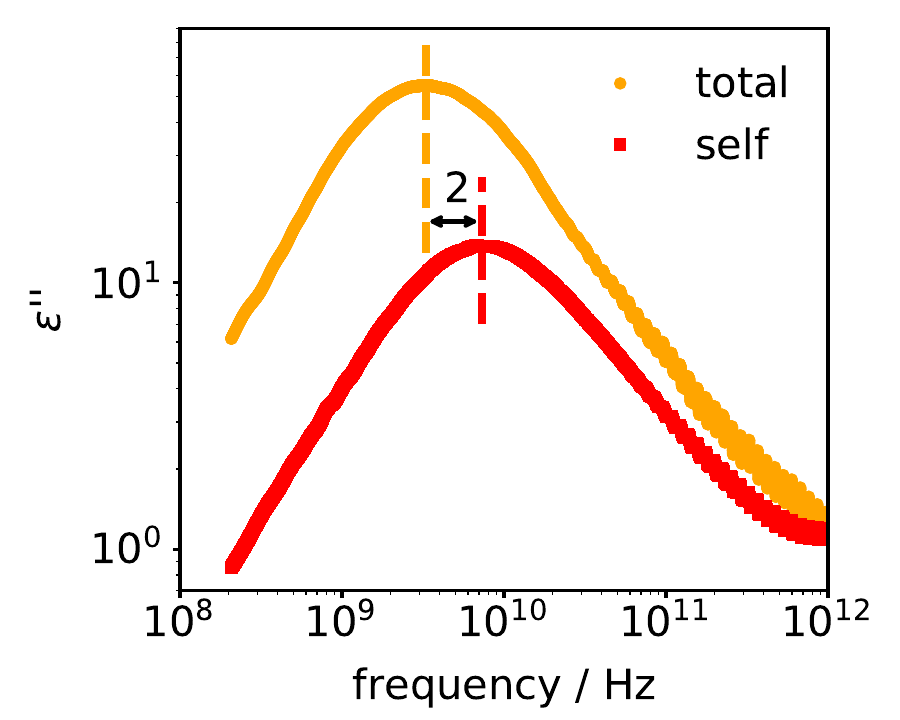}
    \caption{\textbf{Total and Self Dielectric Spectrum} at 260~K. A factor of two between the peak position of total and self spectrum is found, rationalizing the shift of the NMR data (which measures rotational self correlations) by a factor of two to compare with the dielectric time constants.}
    \label{fig:total_self}
\end{figure}

The time constants obtained from nuclear magnetic resonance (NMR) spectroscopy of confined water reported in the literature \cite{steinrucken2024complex} are scaled by a factor of two to facilitate comparison with dielectric relaxation times. This adjustment is necessary because NMR probes the self-correlation contribution of the rotational dynamics, whereas the total dielectric response is dominated by cross-correlations \cite{pabst2025salt}. As shown in Fig.~\ref{fig:total_self}, the peak frequency (and thus the corresponding time constant) differs by approximately a factor of two between the total and self contributions. 
NMR data are included only for temperatures above the melting temperature of water under confinement, in order to exclude any influence of partial crystallization on the measured time constants. Furthermore, the NMR dataset comprises measurements performed predominantly on D$_2$O, with some data obtained for H$_2$O. The isotope effect on the dynamics in this temperature range is small and remains below the differences observed between the two DNN models.

\subsection{Panel B: Pressure dependence}

The pressure-dependent correlation times shown for three temperatures in Fig.~\ref{fig:dyn}B are fitted via an empirical equation of the form
\begin{equation}
   \tau(P) =  a_1\exp(-P/P_0) + a_2 + a_3P^a_4,
   \label{eq:press}
\end{equation}
where $a_1,a_2,a_3,a_4,P_0$ are fitting parameters. The fit is just intended to be a guide to the eye in order to extrapolate the correlation times at 195~K to low pressures, where equilibration is no longer feasible. Already at 2000~bar, the correlation time exceeds the time scales accessible in the simulations. To nevertheless obtain an estimate at this pressure (open symbol in Fig.~\ref{fig:dyn}B), we compute the dielectric spectrum from the partially decayed dipole autocorrelation function, for which the peak position lies outside the accessible frequency range. 

\begin{figure}[h!]
    \centering
    \includegraphics[width=0.5\linewidth]{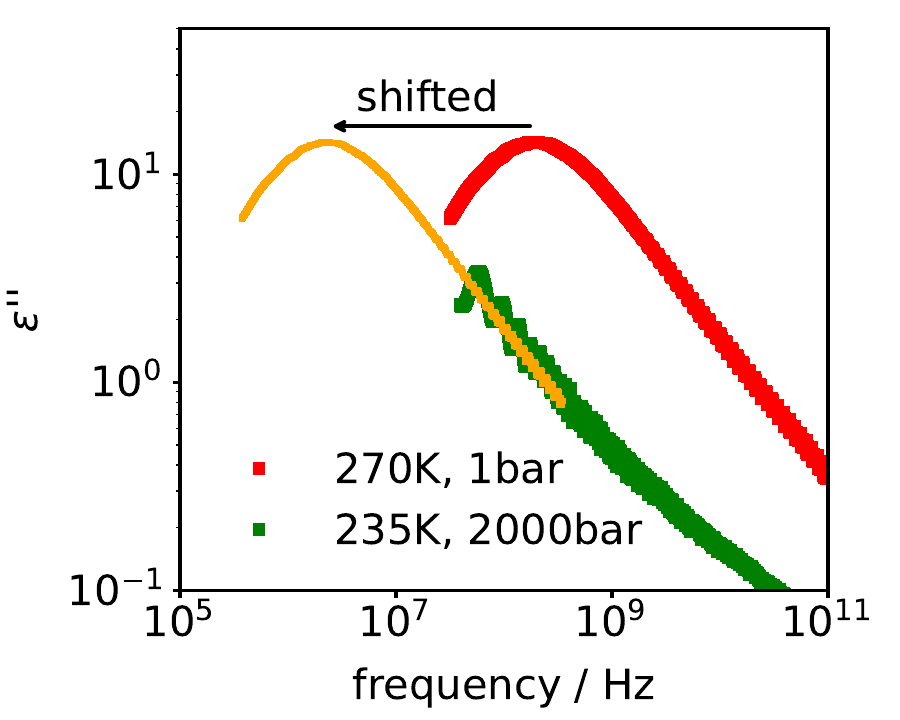}
    \caption{\textbf{Frequency pressure superposition} to estimate the peak position for a simulations where full equilibration can not be reached. This method is only used for the one open circle in Fig.~\ref{fig:dyn}B.}
    \label{fig:extrapol}
\end{figure}

To estimate the peak position, a spectrum obtained from a well equilibrated simulation is shifted to overlap with the partial spectrum at 2000~bar (see Fig.~\ref{fig:extrapol}). It should be noted that the resulting $\tau$ represents a lower bound of the true correlation time, since the simulation is not fully equilibrated, which may bias $\tau$ toward shorter times. The fit with Eq.~\ref{eq:press} illustrates how rapidly $\tau$ varies with pressure at lower temperatures.

\subsection{Panel C: Additional dynamic quantities}
To further justify the choice of the rotational correlation time $\tau$ as a robust measure of structural relaxation in the main text, we benchmark it against other transport properties, demonstrating that $\tau$ remains strongly coupled to the shear viscosity $\eta$, whereas the diffusion coefficient $D$ decouples and becomes comparatively faster at low temperatures, limiting its usefulness for assessing equilibration in the supercooled regime.

\begin{figure}[h!]
    \centering
    \includegraphics[width=0.8\linewidth]{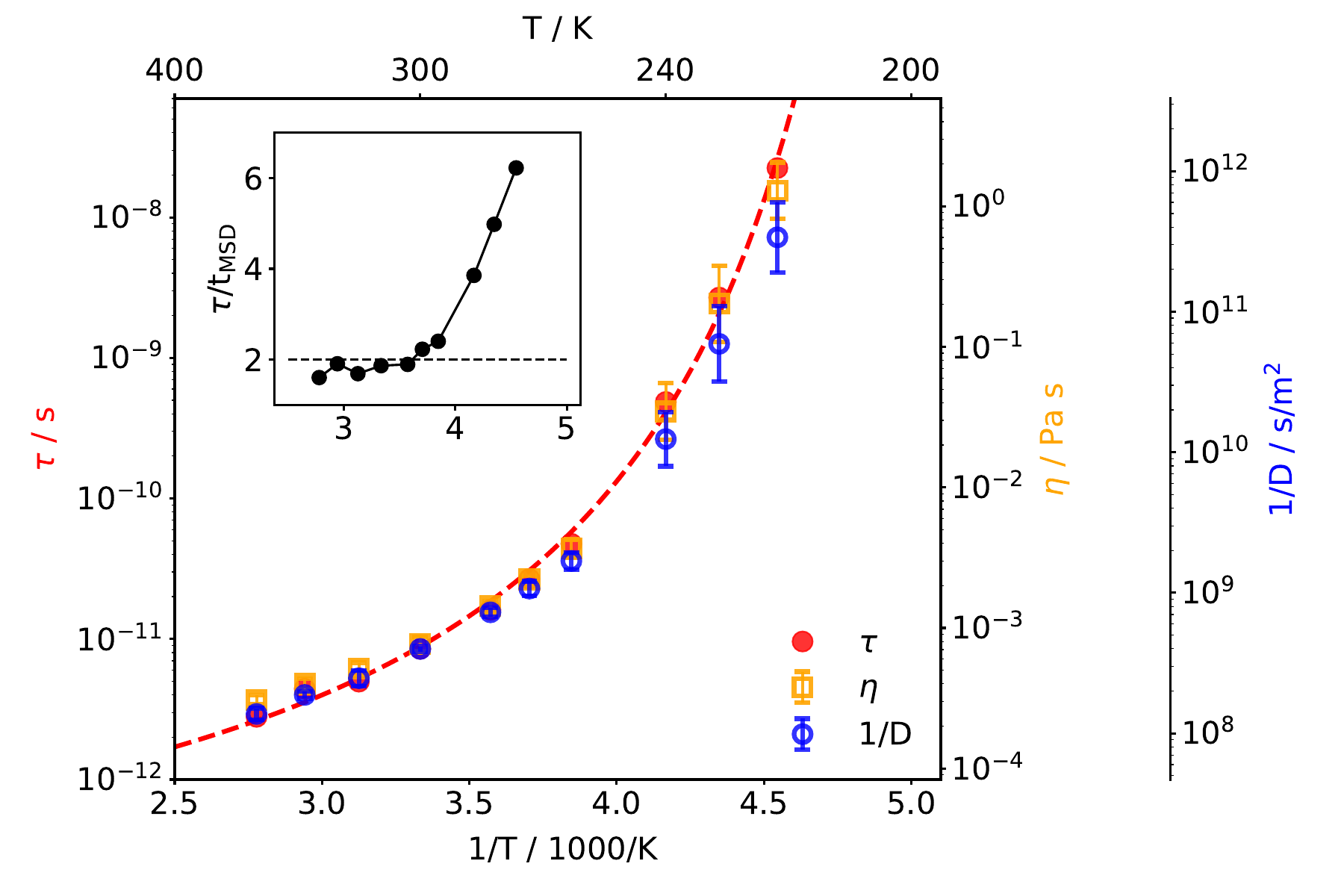}
    \caption{\textbf{Comparison of transport properties.} Relaxation times $\tau$ compared to viscosity and inverse diffusion coefficients. Inset: The ratio $\tau$/$t_{\rm MSD}$, where $t_{\rm MSD}$ is the time for MSD to reach one squared molecular diameter. }
    \label{fig:arrh2}
\end{figure}

In Fig.~\ref{fig:arrh2}, the structural relaxation times $\tau$ obtained with the DNN@SCAN model are compared to the shear viscosity $\eta$ and the inverse diffusion coefficient $D^{-1}$. As expected from experiments \cite{dehaoui2015viscosity}, $\tau \propto \eta$ to a good approximation, while $D$ decouples from $\tau$ and $\eta$ at low temperatures, an effect known as the violation of the Stokes--Einstein relation, and assigned to the jump diffusion mechanism in the case of water \cite{dueby2019decoupling}. The viscosity is calculated via the Green--Kubo formalism from the off-diagonal elements of the pressure tensor using the \texttt{sportran} code \cite{ercole2022sportran}; see Ref.~\cite{pabst2025glassy} for further details. Diffusion coefficients are obtained from the mean square displacement (MSD) via the Einstein relation, with the MSD computed using the \texttt{TRAVIS} trajectory analysis code \cite{brehm2020travis}.

The inset shows the ratio of $\tau$ to the time $t_{\rm MSD}$ required for the MSD to reach the squared molecular diameter (0.28~nm) of a water molecule. In analogy to the diffusion coefficient, a decoupling is observed at low temperatures. At high temperatures, this ratio is approximately equal to two (dashed line), consistent with the common notion that translational diffusion over one molecular diameter corresponds to the rotational self-correlation time, which is approximately a factor of two shorter than $\tau$ (see Sec.~\ref{sec:NMR}).

\subsection{Relation of $T_g$ to other characteristic temperatures.}

Diffusion coefficients extracted from crystal growth measurements have been used to support the notion of a glass transition in water around 136~K, since significant crystal growth (and thus molecular mobility) has been observed in this temperature regime \cite{xu2016growth}. The temperature dependence of the inferred diffusion coefficients follows Arrhenius behavior below $\sim 150$~K and crosses over to a much stronger temperature dependence above $\sim$180~K \cite{xu2016growth}. Such an apparent change in slope is commonly observed when secondary ($\beta$) relaxation processes, which typically exhibit Arrhenius behavior, separate from the structural ($\alpha$) relaxation at $T_g$. 

\begin{figure}[h!]
    \centering
    \includegraphics[width=0.75\linewidth]{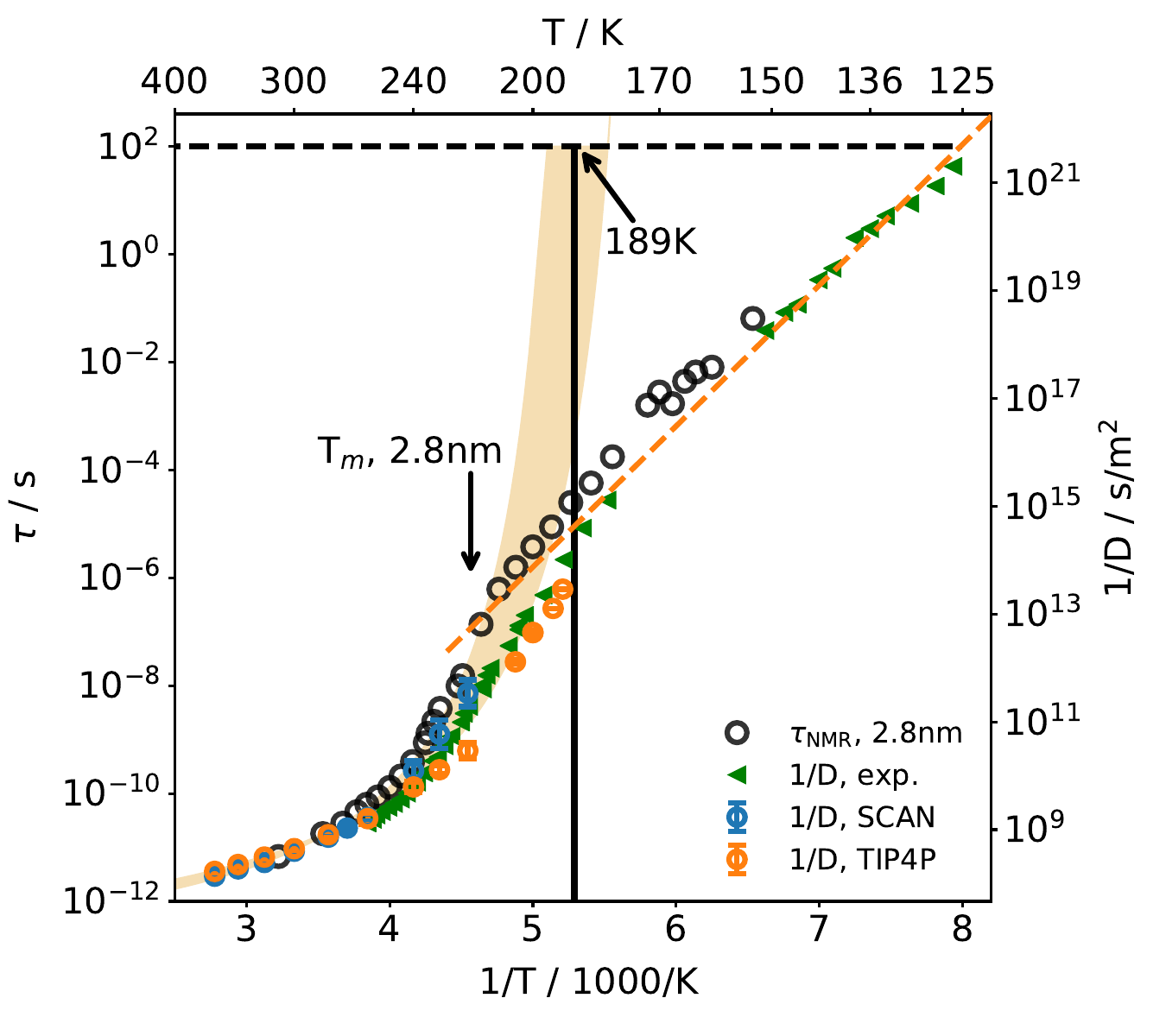}
    \caption{\textbf{Comparison of simulated and experimental diffusion coefficients.} Experimental diffusion coefficients inferred from crystal growth rates are taken from Ref.~\cite{xu2016growth}. Dashed line is an Arrhenius fit to the experimental data below 189~K. We also reproduce the full dataset of NMR relaxation times in 2.8~nm confinement from Ref.~\cite{steinrucken2024complex}, from which we omitted the data below the melting point $T_m$ in Fig.~\ref{fig:dyn}.}
    \label{fig:diffusion}
\end{figure}

Indeed, it is well established for various liquids that crystallization can proceed via molecular mechanisms distinct from the structural relaxation, allowing crystallization to occur even in the glassy state \cite{newman2020we}. To test this scenario, we compare the experimental diffusion coefficients from Ref.~\cite{xu2016growth}, inferred from crystallization rates, with the diffusion coefficients obtained from our simulations in Fig.~\ref{fig:diffusion}. The shaded area denotes the range of structural relaxation times $\tau$ from the different models, as shown in Fig.~\ref{fig:dyn}C. The average glass transition temperature $T_g = 189$~K determined in the main text is indicated by a vertical line. 

The experimental diffusion coefficients below 189~K are well described by an Arrhenius fit. Most importantly, above 189~K, the experimental data deviate from this Arrhenius behavior, reflecting the pronounced change in temperature dependence discussed above. Thus, the experimental diffusion coefficients are fully consistent with a glass transition around $\sim 189$~K, with crystallization below $T_g$ proceeding via secondary relaxation processes rather than structural relaxation. We note that a similar scenario has been proposed for water under confinement \cite{melillo2024complexity}, and we reproduce the full dataset of NMR relaxation times in 2.8~nm confinement from Ref.~\cite{steinrucken2024complex} again shifted by a factor of~2 (see section~\ref{sec:NMR}), from which we omitted the data below the melting point in Fig.~\ref{fig:dyn}. At the melting point in confinement, a kink in the temperature dependence of the NMR relaxation times is observed, closely resembling the one of the diffusion coefficients, underscoring that this behavior might be unrelated to equilibrium dynamics of bulk water.  

However, we also note that the $\beta$-relaxation scenario at the lowest temperatures is challenged by diffusivity measurements on layered H$_2$O-D$_2$O films \cite{kimmel2025translational}.

\begin{figure}[h!]
    \centering
    \includegraphics[width=0.65\linewidth]{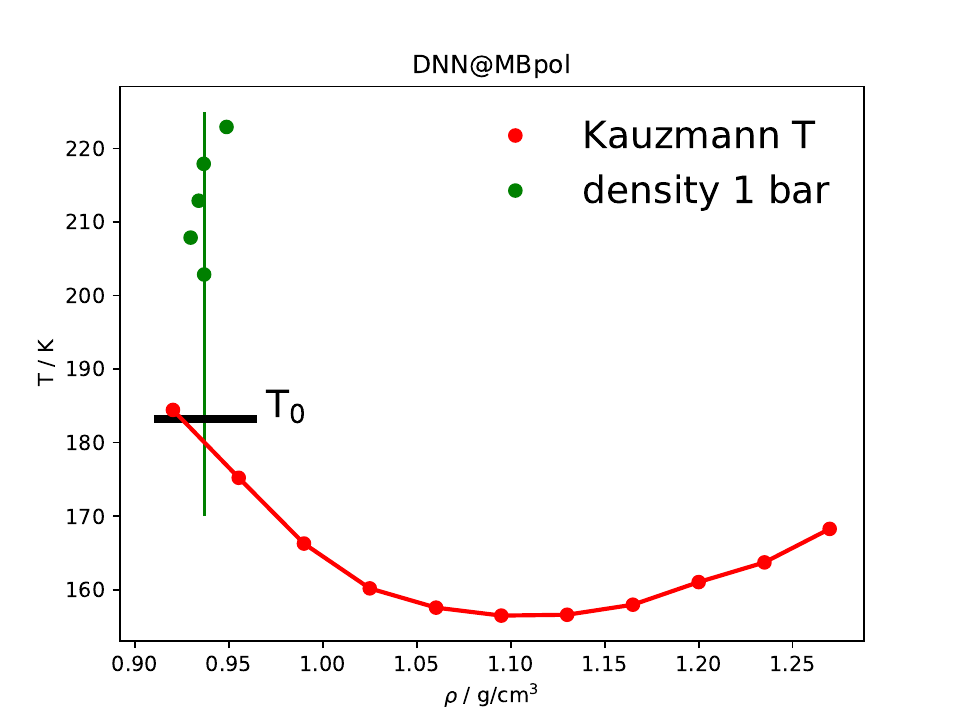}
    \caption{\textbf{Kauzmann temperature of DNN@MB-pol} obtained from extensive energy landscape calculations in \cite{szukalo2026energy}. Density at 1~bar from \cite{sciortino2025constraints}. Our $T_0$ from the VFT fit (black line) is in good agreement with the Kauzman temperature at a density similar to the one obtained at the lowest temperatures accessible.}
    \label{fig:kauz}
\end{figure}

Lastly, we compare our simulated value of $T_g$ with empirical relationships for glass transition temperatures. While these relations are not definitive, they provide useful context. In the main text, we already discussed the coincidence between the experimental estimate of the Kauzmann temperature and our $T_0$ value obtained from the VFT fit, a correspondence observed in many glass-forming liquids. Very recently the Kauzmann temperature was obtained for the DNN@MB-pol model from extensive energy landscape calculations as a function of density \cite{szukalo2026energy}. We reproduce this data together with the density of this model at 1~bar taken from \cite{sciortino2025constraints} as a function of temperature in Fig.~\ref{fig:kauz}. Our $T_0$ from the VFT fit (horizontal black line) is in excellent agreement with the Kauzmann temperature of the model at a density similar to the one obtained at the lowest temperatures accessible. 

Furthermore, many liquids follow the so-called two-thirds rule, according to which the glass transition temperature is approximately two-thirds of the melting temperature \cite{ito1999thermodynamic}. For water, this yields $\frac{2}{3}\cdot 273~\mathrm{K} = 182~\mathrm{K}$, which lies within the uncertainty of our estimate $T_g = 189 \pm 8~\mathrm{K}$. 

In addition, for many systems, fitting correlation times with a power law as in mode-coupling theory yields a divergence temperature $T_c$ that is approximately $1.2\,T_g$ \cite{das2004mode}. For water, using the commonly cited value $T_c = 228~\mathrm{K}$ leads to $T_g = 228~\mathrm{K}/1.2 = 190~\mathrm{K}$, again in excellent agreement with our estimate. 

Taken together, these observations indicate that, from the perspective of glass-forming liquids, water at ambient pressure does not exhibit anomalous behavior but rather conforms to established empirical trends, displaying VFT-like structural relaxation and the emergence of secondary relaxation processes at the glass transition temperature.

\section{Supplementary Information for Fig.~4}\label{sec:SI-density}

In order to obtain the inter-/extrapolated densities along the isochrones in Fig.~\ref{fig:phase}, we fitted the densities from simulations at different temperatures and pressures with a polynomial of the form

\begin{equation}
    \rho(T,P) = a_1 T + a_2 P + a_3 T^2 + a_4 TP + a_5 P^2.
    \label{eq:poly}
\end{equation}

\begin{figure}[h!]
    \centering
    \includegraphics[width=0.5\linewidth]{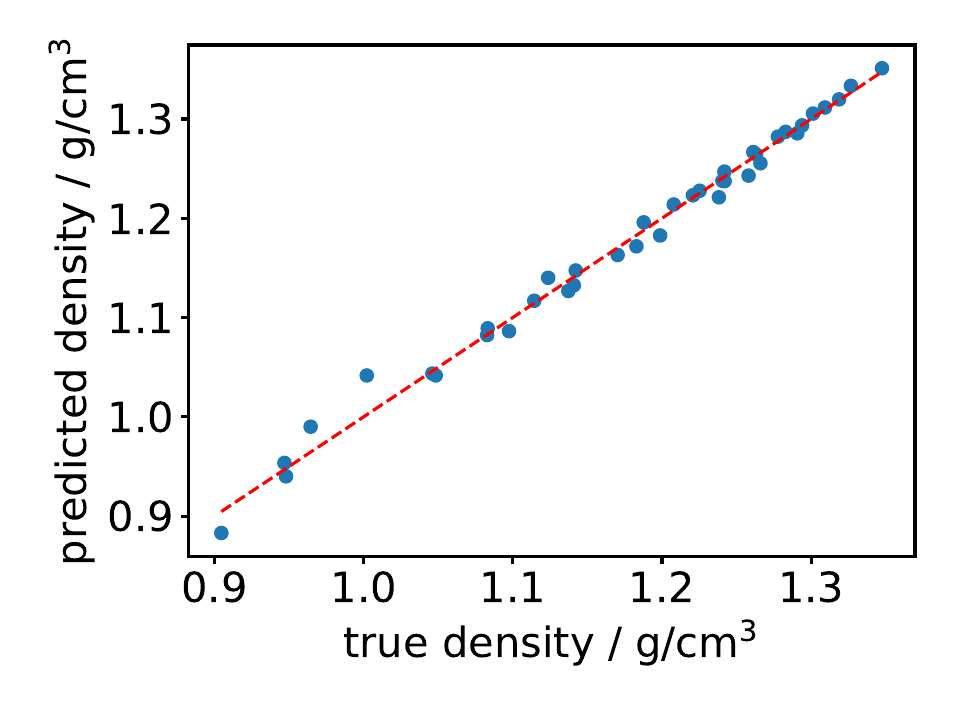}
    \caption{Densities from simulations at different temperatures and pressures compared to the ones predicted by the polynomial fit (eq.~\ref{eq:poly}).  }
    \label{fig:density}
\end{figure}

The comparison between the true densities and the ones predicted by the polynomial fit is shown in Fig.~\ref{fig:density}.

%



\end{document}